\documentclass[]{aa}
\usepackage{graphicx}
\usepackage{natbib} \bibpunct{(}{)}{;}{a}{}{,}
\usepackage{color}
\usepackage[varg]{txfonts}
\usepackage{rotating}
\def\changed{}
\def\changedA{}

  \newcommand{\msunpyr}{M_\odot\,\mbox{yr}^{-1}}
 \newcommand{\dint}{\,\mbox{d}} \newcommand{\ddif}{\mbox{d}}
\newcommand{\kms}{\ifmmode{\,\mbox{km}\,\mbox{s}^{-1}}\else{km/s}\fi}
\newcommand{\msun}{\ifmmode M_{\odot} \else M$_{\odot}$\fi}
\newcommand{\rsun}{\ifmmode R_{\odot} \else R$_{\odot}$\fi}
\newcommand{\lsun}{\ifmmode L_{\odot} \else L$_{\odot}$\fi}
\newcommand{\zsun}{\ifmmode Z_{\odot} \else $Z_{\odot}$\fi}
\newcommand{\velo}{\ifmmode\varv\else$\varv$\fi}
\newcommand{\vinf}{\ifmmode\velo_\infty\else$\velo_\infty$\fi}
\begin{document} 

\title{Stellar mass-loss near the Eddington limit} \subtitle{Tracing
  the sub-photospheric layers of classical Wolf-Rayet stars}
 
%\titlerunning{}
 
\author{G.\ Gr\"{a}fener\inst{\ref{inst1}} \and J.S.\
  Vink\inst{\ref{inst1}}}

\institute{Armagh Observatory, College Hill, Armagh, BT61\,9DG, United
  Kingdom\label{inst1}}

\date{Received ; Accepted}
 
%\abstract
%%
%{Context}
%%
%{Aims}
%%
%{Methods}
%%
%{Results}
%%
%{Conclusions}

\abstract{Towards the end of their evolution, hot massive stars
  develop strong stellar winds and appear as emission line stars, such
  as Wolf-Rayet (WR) stars or luminous blue variables (LBVs). The
  quantitative description of the mass loss in these important
  pre-supernova phases is hampered by unknowns, such as clumping and
  porosity due to an inhomogeneous wind structure and by an incomplete
  theoretical understanding of optically thick stellar winds. Even the
  stellar radii in these phases are badly understood since they are
  often variable (LBVs) or deviate from theoretical expectations (WR
  stars).}  {In this work we investigate the conditions in deep
  atmospheric layers of WR stars to find out whether they comply with
  the theory of optically thick winds and whether we find indications
  of clumping in these layers.}{We used a new semi-empirical method to
  determine sonic-point optical depths, densities, and temperatures
  for a large sample of WR stars of the carbon (WC) and oxygen (WO)
  sequence.  Based on an artificial model sequence we investigated the
  reliability of our method and its sensitivity to uncertainties in
  stellar parameters.}  {We find that the WR stars in our sample obey
  an approximate relation with $P_{\rm rad}/P_{\rm gas} \approx 80$ at
  the sonic point. This `wind condition' is ubiquitous for radiatively
  driven, optically thick winds, and it sets constraints on possible
  wind/envelope solutions affecting radii, mass-loss rates, and
  clumping properties.}  {Our results suggest that the presence of an
  optically thick wind may force many stars near the Eddington limit
  to develop clumped, radially extended sub-surface zones.  The
  clumping in these zones is most likely sustained by the non-linear
  strange-mode instability and may be the origin of the observed wind
  clumping.  The properties of typical late-type WC stars comply with
  this model.  Solutions without sub-surface clumping and inflation
  are also possible but require compact stars with comparatively low
  mass-loss rates.  These objects may resemble the small group of WO
  stars with their exceptionally hot stellar temperatures and highly
  ionized winds.}

\maketitle

\section{Introduction} 
\label{sec:intro} 

Mass loss through radiatively driven stellar winds is crucial for the
evolution of massive luminous stars. One aspect is the direct removal
of mass from the outer layers and subsequent exposition of chemically
enriched material on the stellar surface \citep{con1:76}. Depending on
the environment metallicity, this can affect the final core masses of
massive stars before supernova (SN) explosion and thus dictate how
massive stars end their lives \citep{heg1:03}. In line with the
removal of mass, angular momentum is efficiently removed from the
stellar surface. This can affect the rotational properties of massive
stars \citep[e.g.][]{vin1:10} and, again depending on metallicity,
determine whether they end their lives as ordinary SNe or gamma-ray
bursts (GRBs) \citep[e.g.][]{pet1:05,yoo1:05,gra2:12}.

In the present work we are particularly interested in phases of
exceptionally strong mass loss, as it occurs in the Wolf-Rayet (WR)
and luminous blue variable (LBV) phases. We use a new method to
determine the conditions near the sonic point of optically thick
stellar winds, thus gaining information on the otherwise unobservable
sub-surface layers of these objects. In Sect.\,\ref{sec:physics} we
give a detailed introduction to the underlying concepts of this work.
In Sect.\,\ref{sec:theory} we discuss the theoretical background of
our method, and validate its applicability using fully self-consistent
wind models for WR stars. In Sect.\,\ref{sec:application} we apply our
method to a large sample of Galactic WC stars. We employ two
approaches that differ with respect to the amount of empirical vs.\
theoretical input. In Sect.\,\ref{sec:sequence} we construct an
artificial model sequence for WC stars to investigate the sensitivity
of our results to uncertainties in the stellar parameters, and the
general applicability of our results. In Sect.\,\ref{sec:discussion}
we discuss the consequences of our results for the physics of
optically thick winds and their sub-photospheric layers. Conclusions
are drawn in Sect.\,\ref{sec:conclusions}.

\section{The envelopes and winds of stars near the Eddington limit}
\label{sec:physics}

In the WR and LBV phases massive stars develop strong stellar winds,
most likely owing to their proximity to Eddington limit.  The wind
densities in these phases are so high that the winds become optically
thick and develop characteristic emission-line spectra (cf.\
Sect.\,\ref{sec:wrmassloss}).  The influence of density
inhomogeneities (in the following referred to as `clumping') plays a
key role in these phases.  On the one hand, clumping affects the
mass-loss diagnostics, and thus introduces significant uncertainties
to empirical mass-loss estimates (cf.\ Sect.\,\ref{sec:clumping}).  On
the other hand, clumping may affect the stellar envelope structure
directly below the surface, leading to a radius inflation by several
factors (cf.\ Sect.\,\ref{sec:inflation}). This effect may be related
to the so-called `radius problem' of WR stars, namely that the
empirically determined radii of WR stars exceed theoretical
expectations by up to a factor 10.  Possible explanations for this
problem are the formation of pseudo-photospheres at large radii within
the stellar winds, or radially extended sub-surface layers due to the
inflation effect (cf.\ Sect.\,\ref{sec:radii}).

An investigation of the conditions in deep atmospheric layers near the
sonic point can help distinguish between these two scenarios and
provide important information about the origin of wind clumping and
its role in the physics of optically thick winds. In the present work
we estimate sonic-point temperatures and densities for a large sample
of WR stars, based on optical depths inferred from their observable
wind momenta (cf.\,Sect.\,\ref{sec:tauwr}).

\subsection{Mass loss near the Eddington limit}
\label{sec:wrmassloss}

Hot stars in late evolutionary stages often show emission line spectra
characteristic for strong stellar winds. This holds for massive
WR stars and LBVs as well as
for their low-mass counterparts, the WR-type central stars of
planetary nebulae ([WR]-CSPNe). Although both types of stars result
from totally different evolutionary sequences and have different
internal structures, they share common properties, namely their
surface enrichment with the products of He-burning (for WC spectral
types) or the CNO cycle (for WN spectral types), and their proximity
to the Eddington limit (i.e., high $L/M$ ratios of the order of $10^4
L_\odot/M_\odot$).

As the strong mass-loss of massive WR stars can substantially affect
the stellar evolution in direct pre-supernova phases, it is desirable
to understand the underlying physical mechanisms that are responsible
for its occurrence.
Very massive stars above $\sim 100\,M_\odot$ may provide the key to
answer this question. Spectral analyses of H-rich WN stars in the
Galaxy \citep{ham1:06,mar1:08} and the LMC \citep{cro1:10,bes1:11}
imply that these objects are very massive main-sequence stars with
high $L/M$ ratios, but partly even solar-like hydrogen abundances.
\citet{gra1:11} found empirical evidence that the mass-loss properties
of the most massive stars in the Arches cluster can be described by a
dependence on the Eddington factor $\Gamma_{\rm e}$.  This result is
in line with theoretical studies indicating that optically thick
WR-type winds can be driven by radiation
\citep{luc1:93,gra1:05,vin1:05} and may be triggered by the proximity
to the Eddington limit \citep{gra1:08,vin1:11}. The last two studies
also support the view by \citet{nug1:02} that the conditions at the
sonic point in the deep, optically thick layers of WR-type winds are
crucial for this kind of mass-loss.

\subsection{Wind clumping}
\label{sec:clumping}

Observationally, the quantitative analysis of WR-type winds is
hampered by the presence of wind-inhomogeneities, or `wind clumping'.
\citet{ham1:98} showed that the electron scattering wings of strong WR
emission lines are generally weaker than predicted by smooth wind
models.  This indicates that the true electron densities in WR winds
are lower than assumed in smooth wind models.  The discrepancy can be
resolved by introducing a wind clumping factor $D$ which leads to an
increased density $n = D \times \bar{n}$ within clumps, and a lower
mean wind density $\bar{n} = n/D$.  For recombination lines, the
line emissivity per volume then scales with $j \propto n^2$, and the
spatial mean with $\bar{j} \propto n^2/D = \bar{n}^2 \times D$.  The
(mean) wind densities determined in spectral analyses thus scale with
$\bar{n} \propto \sqrt{\bar{j} /D}$, i.e.\ empirically determined
mass-loss rates are reduced by a factor $\sqrt{D}$.
For typical clumping factors of the order of 10, WR mass-loss
rates thus have to be reduced by a factor $\sqrt{D} \sim 3$.

For OB stars even higher mass-loss reductions by a factor of
$\sim$\,10 have been proposed, based on the {\changedA weakness of the
  P-Cygni type absorption troughs of some trace element ions
  \citep[e.g.][]{ful1:06}. \citet{osk1:07} could, however, show that
  this effect may be caused by a porous wind structure with optically
  thick clumps. In this case photons may be shielded or leak through
  gaps in between clumps. Dependent on the geometry and size
  distribution of the clumps the mean opacity may then be effectively
  reduced with respect to the non-porous case and the detailed line
  shapes can be explained with moderate clumping factors \citep[cf.\
  also][]{sun1:10,sun1:11}.}
  As changes in $\dot{M}$ of the order of 3--10 are
relevant for stellar evolution it is important to gain detailed
insight in the processes that lead to wind clumping, and to find
additional diagnostics for its presence or absence.

For WR stars the density diagnostics via the strength of the electron
scattering wings is comparatively direct. Nevertheless the precision
of the determined clumping factors $D$ is only moderate, and the
clumping diagnostics is restricted to the formation region of the
scattering wings. It is thus not entirely clear if clumping is only
important in the outer wind, or if it also plays a role in deeper wind
layers which cannot be directly observed. In these layers clumping may have
an effect on the mass-loss rates, and on the spatial extension of the
sub-surface layers of WR stars.

\subsection{Sub-photospheric envelope inflation and clumping}
\label{sec:inflation}

It is a long-standing problem that the radii of WR stars as determined
in spectral analyses are larger than predicted. This is particularly
the case for classical hydrogen-free WR stars which are expected to lie on the
helium main-sequence, i.e.\ at stellar temperatures $T_\star\gtrsim
100$\,kK\footnote{Here $T_\star$ denotes the effective temperature
  related to the radius $R_\star$ near the hydrostatic wind base at
  large optical depth, i.e.\ $T_\star$ is defined via the equation
  $L=4 \pi R_\star^2 \sigma T_\star^4$.}. The temperatures obtained
from spectral analyses, however, go down to $T_\star \sim 30$\,kK, i.e.\
the radii of hydrogen-free WR stars are up to a factor 10 larger than
predicted \citep[e.g.][]{ham1:06,san1:12}.

A common explanation for this effect is the formation of
pseudo-photospheres due to the large spatial extension of optically
thick winds.  Due to this effect the photospheres of WR stars and LBVs
can be located at radii much larger than the hydrostatic stellar
radius $R_\star$ \citep[e.g.][]{del1:82,smi1:04}.  \citet{ham1:04}
have shown that this effect is of major importance for the WR stars
with the strongest mass-loss rates.  These objects lie in a parameter
range where it is only possible to give upper limits for their radii.
For WR stars with moderate mass-loss rates this argument is, however,
difficult to sustain. The atmosphere models used in current spectral
analyses of WR stars take the wind extension fully into account. At
least in a first-order approximation the obtained stellar radii and
temperatures are thus likely correct. Moderate uncertainties may,
however, arise from deviations from the adopted velocity structure in
sub-photospheric layers that are not directly observed \citep[cf.\ the
discussion in][]{gra1:12}.

{\changedA An alternative explanation has been provided by
  \citet{gra1:12} who could explain the increased WR radii by an
  inflation of the outer stellar envelope due to the Fe-opacity peak
  at temperatures around 150\,kK. A quantitative agreement with
  observed WR radii could, however, only be obtained if an increased
  mean opacity within the inflated zone was adopted.  This was
  achieved by assuming an inhomogeneous density structure within the
  inflated zone, where the Rosseland mean opacity is enhanced due to
  the higher density within clumps.}

The inflation effect is expected to occur near the Eddington limit,
and leads to the formation of a low-density sub-surface layer that is
mainly supported by radiation pressure with a density inversion on top
of it.  The resulting radiation-dominated cavities are known to be
instable with respect to linear strange-mode pulsations
\citep{sai1:98,gla1:02}. {\changedA Furthermore \citet{sha2:01} found
  two types of instabilities that become dynamically important near
  the Eddington limit.}  For the non-linear regime \citet{gla1:08}
predicted the formation of density inhomogeneities in the form of
shocks combined with a radial inflation, in line with the results by
\citet{gra1:12}.

In this - theoretically consistent - picture clumping thus develops
from an instability in deep layers near the Fe-opacity peak. As a
consequence the layers around the opacity peak inflate. For the theory
of optically thick winds this means that the effects of
inhomogeneities need to be taken into account. In particular, the mean
opacity may be enhanced near the sonic point. {\changedA At this point
  the theory of optically thick winds \citep[e.g.][]{pis1:95,nug1:02}
  imposes a critical condition where the radiative acceleration
  $g_{\rm rad}$ approximately equals gravity, i.e.\ the Eddington
  factor $\Gamma = g_{\rm rad}/g \approx 1$.  In addition to this,}
the radii of WR stars cannot be seen as a fixed quantity anymore.
They may adjust to the boundary conditions imposed by the stellar wind
\citep[cf.\ Sect.\ 3.5.1 in][]{gra1:12}, adding further complexity to
the question how WR winds form.

{\changedA In this context it is important to note that, {\changed
    analogous to the situation in stellar winds (cf.\
    Sect.\,\ref{sec:clumping}), porosity may reduce the mean opacity
    in the sub-surface layers, and thus counteract the effect of
    clumping.} For this to happen it is necessary that the geometry
  deviates significantly from a shell structure (as e.g.\ in 1D
  simulations) so that photons can leak through gaps between clumps.
  Conditions like this have been investigated mainly for stellar
  interiors and winds above the Eddington limit and may also affect
  the dynamics of WR winds \citep{sha1:98,sha1:01,owo1:04,mar2:09}.
  As a result of porosity the true clumping factors within the
  inflated zone may be higher than the values adopted by
  \citet{gra1:12}.}

Based on dynamical models, \citet{pet1:06} found that an envelope
inflation may be totally inhibited by strong stellar winds. The
limiting mass-loss rate depends on mass and radius of the star, and it
is not entirely clear whether an envelope inflation is generally
inhibited for WR stars or not \citep[cf.\ the discussion
in][]{gra1:12}.  We note that the limiting mass-loss rate increases
with radius, and plays no significant role for cooler stars, such as
LBVs.

\subsection{Wind extension vs.\ envelope inflation}
\label{sec:radii}

Following the discussion in Sect.\,\ref{sec:inflation} we distinguish
between two possible scenarios to explain the observed radii of
hydrogen-free WR stars. These are 1) wind extension, and 2) envelope
inflation. A criterium to distinguish between the two scenarios is the
location of the sonic radius $R_{\rm s}$. By definition $R_{\rm s}$ is
the radius where dynamic terms start to dominate the equation of
motion, i.e.\ below $R_{\rm s}$ the stellar atmosphere is in a
quasi-static equilibrium while above $R_{\rm s}$ we have a dynamically
flowing wind.  The density structure below $R_{\rm s}$ thus follows an
exponential distribution, leading to a rapid increase in the optical
depth below $R_{\rm s}$. For this reason $R_{\rm s}$ almost coincides
with the `hydrostatic' stellar radius $R_\star$ that is determined in
spectral analyses, i.e.\ we have $R_{\rm s}\approx R_\star$.

In scenario 1) $R_{\rm s}$ (and thus also $R_\star$) is small, and
$T_\star$ is high ($\gtrsim 100$\,kK).  To explain the low observed
$T_\star$, the spatial extension of the super-sonic layers {\em above}
$R_{\rm s}$ has to be larger than assumed in the atmosphere models.
To achieve this the real density and velocity structure has to deviate
significantly from the $\beta$-type velocity laws
(Eq.\,\ref{eq:betalaw}) that are usually adopted in the models.  In
this scenario clumping may be present near the sonic point, but it is
not a mandatory condition to launch the WR wind.

An example of such a case is the self-consistent hydrodynamical model
by \citet{gra1:05} with $T_\star=140$\,kK.  In this model the inner
part of the wind is driven by Fe M-shell ions (Fe\,{\sc ix--xvi}),
which are exactly the ions responsible for the Fe-opacity peak near
150\,kK.  While these opacities provide the wind acceleration near the
sonic point and slightly above, the outer wind is accelerated by a
`cool' opacity bump due to lower ionization stages.  Between the two
opacity bumps the wind has to cross a region of reduced mean opacity,
and forms a velocity plateau. If such a plateau occurs below the
photosphere of an optically thick wind it can mimic a star with lower
$T_\star$ and a `standard' velocity structure.  E.g., the hydrodynamic
model by \citet{gra1:05} matches the spectral appearance of the
galactic WC star WR\,111. Based on models with a $\beta$-type velocity
structure \citet{gra1:02} determined $T_\star=85$\,kK for the same
object. We note here that the model by \citet{gra1:05} already
represents a somewhat extreme case, and that the wind density in this
model is high. For WR stars with lower wind densities and even lower
observed $T_\star$ it may be difficult to employ this scenario.

In scenario 2) $R_{\rm s}$ (and thus also $R_\star$) is located at
much larger radii due to an inflation of the sub-surface layers below
$R_{\rm s}$. Consequently $T_\star$ is lower than predicted by
standard stellar structure models. \citet{gra1:12} have shown that
clumping is mandatory do achieve such an envelope inflation for WR
stars in the observed parameter range.  In this scenario the
sub-surface layers near the Fe-opacity peak are clumped and the
observed clumping in the winds of WR stars may originate from these
layers.  In particular, clumping would most likely be present at the
sonic radius $R_{\rm s}$ and affect the critical-point conditions of
WR winds.

\subsection{WR wind momentum and sonic-point conditions}
\label{sec:tauwr}

The goal of the present work is to investigate the conditions near the
sonic radius $R_{\rm s}$, mainly to find out if these comply with the
present `smooth' wind theory for WR stars \citep[e.g.][]{nug1:02}, and
to obtain information on clumping and inflation of sub-surface layers,
in line with our discussion in Sect.\,\ref{sec:radii}.

To this purpose we employ a relation between wind efficiency and
optical depth, namely that $\eta = \dot{M}\varv_\infty/(L/c) \approx
\tau_{\rm s}$ (cf.\ Sect.\,\ref{sec:taus}).
{\changed Here we use the formulation from
  \citet[][Sect.\,7.2]{lam1:99} which is based on the sonic-point
  optical depth $\tau_{\rm s}$ \citep[in contrast to similar works
  by][]{net1:93,gay2:95}.  This relation has recently been applied by}
\citet{vin1:12} to calibrate the mass-loss rates of very massive stars
with $\eta=\tau=1$. We note that this relation is independent of wind
clumping and porosity which makes it a powerful tool to address the
problems discussed in Sect.\,\ref{sec:clumping}.

The above relation is of fundamental importance for WR stars and many
LBVs. Both types of stars tend to show strong emission-line spectra
that are a sign of a high wind optical depth $\tau_{\rm s}$.
The emission lines are caused by recombination cascades that occur
when major constituents of the wind material (such as H or He for WN
stars and LBVs, or He, C, and O for WC stars) start to change their
state of ionization. Because the dominant ionization source is
photoionization, this means that the optical depth of the wind
material must be large enough to absorb the majority of ionizing
photons within the wind itself, i.e.\ the wind optical depth must be
high.

This requirement of a large $\tau_{\rm s}$ is intimately linked to the
so-called `wind momentum problem', namely that the wind momenta
$\dot{M}\varv_\infty$ of WR stars and LBVs exceed the momentum
provided by their radiation field $L/c$. This has often been used as
an argument against radiative driving as the source of wind
acceleration for these objects. In fact, however, a large wind optical
depth implies that photons are absorbed and re-emitted more than once
{\em within the wind itself} before they escape the stellar wind,
i.e.\ radiative driving {\em has to lead} to a wind efficiency $\eta =
\dot{M}\varv_\infty/(L/c) > 1$.

In the remainder of this work we take advantage of the fact that
$\eta$, {\changedA as determined in previous spectral analyses,} gives
us clumping-independent information about $\tau_{\rm s}$, and thus
also about temperature $T_{\rm s}$ and density $\rho_{\rm s}$ near the
sonic point.

\section{Theoretical background}
\label{sec:theory}

In this section we describe how we estimate the sonic point
temperatures and densities, $T_{\rm s}$ and $\rho_{\rm s}$, from the
observed wind efficiencies $\eta = \dot{M}\varv_\infty/(L/c)$ of stars
with radiatively driven, optically thick winds. In
Sect.\,\ref{sec:taus} we describe the underlying relation $\eta
\approx \tau_{\rm s}$ in analogy to \citet[][]{lam1:99}.  In
Sect.\,\ref{sec:trel} we obtain $T_{\rm s}$ and $\rho_{\rm s}$ based
on a relation by \citet{luc1:71,luc1:76,luc1:93}. In
Sect.\,\ref{sec:tests} we perform a direct comparison with the
temperature structure of a hydrodynamic atmosphere/wind model computed
in non-LTE, to verify our approach.  Finally, in
Sect.\,\ref{sec:beta}, we establish a simplified approach assuming a
$\beta$-type velocity law to obtain $T_{\rm s}$ and $\rho_{\rm s}$ in
the general case.

\subsection{Estimating the wind optical depth $\tau_{\rm s}$}
\label{sec:taus}

In the following we assume that the winds of WR stars are radiatively
driven. In this case the underlying equations for the dynamics of a
radially expanding stationary stellar wind are the equation of motion
\begin{equation}
\label{eq:motion}
\rho \varv \frac{\ddif \varv}{\ddif r}
= - \frac{\ddif P_{\rm gas}}{\ddif r} - \frac{\ddif P_{\rm rad}}{\ddif r} - \rho g,
\end{equation}
and the equation of continuity
\begin{equation}
\label{eq:cont}
\dot{M} = 4\pi \rho \varv r^2.
\end{equation}
Here $P_{\rm gas}$ and $P_{\rm rad}$ are the gas pressure and
radiation pressure, and $g$ is the local gravitational acceleration
$GM/r^2$. The gradient of $P_{\rm rad}$ in Eq.\,\ref{eq:motion} is
related to the (outward directed) radiative acceleration $g_{\rm rad}$
via
\begin{equation}
\label{eq:prad}
\frac{\ddif P_{\rm rad}}{\ddif r} = - \rho \kappa_F \frac{F_{\rm rad}}{c}
 = - \rho \kappa_F \frac{L}{4\pi r^2 c} = - \rho g_{\rm rad}.
\end{equation}
Here $F_{\rm rad}$ is the frequency-integrated radiative flux
\begin{equation}
\label{eq:flux}
  F_{\rm rad} = \int_0^\infty F_\nu \dint\nu
\end{equation}
and $\kappa_F$ the flux-weighted mean opacity
\begin{equation}
\label{eq:flux-mean}
\kappa_F = \frac{1}{F_{\rm rad}} \int_0^\infty \kappa_\nu F_\nu \dint\nu.
\end{equation}

It is important to note that the above equations describe a smooth and
stationary gas flow. However, in reality stellar winds are believed to
be structured. In this case $\rho$ may indicate a mean density, and
$\kappa_F$ a (spatial) mean opacity that includes effects like
clumpiness and porosity of the wind material
\citep[cf.][]{ham1:98,osk1:07}. We wish to emphasize that the
following derivations are still valid in this case, i.e.\ we assume
that $\kappa_F$ includes effects like clumping and porosity.

To estimate the sonic-point optical depth $\tau_{\rm s}$ we make use
of the fact that, due to hydrostatic equilibrium in the subsonic
region (i.e.\ for $r<R_{\rm s}$), the terms on the right hand side of
Eq.\,\ref{eq:motion} cancel each other and become essentially zero. In
the supersonic region the situation is different as radiation pressure
starts to dominate the dynamics of the gas flow.  In this case $P_{\rm
  gas}$ can be neglected and after multiplication with $4\pi r^2$
Eq.\,\ref{eq:motion} becomes
\begin{equation}
  4\pi \rho \varv r^2 \dint \varv = 4\pi r^2 \rho (g_{\rm rad}-g) \dint r.
\end{equation}
With the equation of continuity (Eq.\,\ref{eq:cont}), and the Eddington factor 
$\Gamma=g_{\rm rad}/g$ this equation
becomes
\begin{equation}
  \dot{M} \dint \varv = 4\pi G M \rho (\Gamma-1) \dint r.
\end{equation}
Using the definitions in Eq.\,\ref{eq:prad} we obtain for $r>R_{\rm
  s}$
\begin{equation}
  \frac{\dot{M}}{L/c} \dint \varv = \kappa_F \rho \frac{\Gamma-1}{\Gamma} \dint r = \frac{\Gamma-1}{\Gamma} \dint \tau
\end{equation}
with the flux-mean optical depth $\tau$. Taking into account that the
left hand side of Eq.\,\ref{eq:motion} becomes small due to
hydrostatic equilibrium below $R_{\rm s}$, the integral of
(Eq.\,\ref{eq:motion}) $\times 4\pi r^2$ becomes
\begin{equation}
  \int_0^{\varv_\infty} \frac{\dot{M}}{L/c}  \dint \varv =  \frac{\dot{M}\varv_\infty}{L/c} 
  = \int_{R_{\rm s}}^\infty  \frac{\Gamma-1}{\Gamma}  \dint \tau
  \approx \tau_{\rm s}.
\end{equation}
In the last step we assumed that $\Gamma$ is significantly larger than
one in the supersonic region. In reality $\Gamma$ will, however, only be
moderately larger than one, so that we expect that
\begin{equation}
\label{eq:taus}
\eta =  \frac{\dot{M}\varv_\infty}{L/c} = f \tau_{\rm s},
\end{equation}
with $f \lesssim 1$ \citep[cf.][]{vin1:12}. We can thus gain
information about $\tau_{\rm s}$ from the basic stellar/wind
parameters $\dot{M}$, $\varv_\infty$, and $L$ as they are routinely
determined in spectral analyses.

Under optically thick conditions, i.e.\ for large $\tau_{\rm s}$, the
temperature is connected to $\tau$ via the radiative diffusion
equation. We can thus gain important information about the sonic-point
temperature $T_{\rm s}$ in otherwise unobservable layers below the
photosphere, just by investigating basic stellar/wind parameters.

\subsection{The temperature structure in optically thick winds}
\label{sec:trel}

To estimate the temperature structure $T(r)$ in the sub-photospheric
layers of an optically thick wind we compute the radiation pressure
$P_{\rm rad}(r)$ from Eq.\,\ref{eq:prad}.  We assume that the
radiation pressure $P_{\rm ref}$ at a reference radius $R_{\rm ref}$
is known, and rewrite Eq.\,\ref{eq:prad} in terms of the radiative
flux $F_{\rm ref}$ at this radius.
\begin{equation}
\label{eq:dprad}
\ddif P_{\rm rad} = -\rho\kappa_F \frac{F_{\rm rad}}{c} \ddif r
= - \rho \kappa_F \frac{F_{\rm ref}}{c} \frac{R_{\rm ref}^2}{r^2} \ddif r 
= \frac{F_{\rm ref}}{c} \ddif \tilde{\tau},
\end{equation}
with the modified optical depth $\tilde{\tau}$ defined by
\begin{equation}
\label{eq:taumod}
\ddif \tilde{\tau} 
= - \rho \kappa_F \frac{R_{\rm ref}^2}{r^2} \ddif r.
\end{equation}
$F_{\rm ref}$ can be computed from the stellar luminosity $L$ and
$R_{\rm ref}$. We can define an effective temperature (or flux
temperature) $T_{\rm ref}$ related to this radius
\begin{equation}
  \label{eq:tref}
  F_{\rm ref} = \frac{L}{4\pi R_{\rm ref}^2} = \sigma T_{\rm ref}^4.
\end{equation}
With these definitions we can compute $P_{\rm rad}$ directly from
Eq.\,\ref{eq:dprad}
\begin{equation}
\label{eq:pradf}
P_{\rm rad} = P_{\rm ref} + \frac{F_{\rm ref}}{c} \left(\tilde{\tau} - \tilde{\tau}_{\rm ref}\right).
\end{equation}
This result is exact if the modified optical depth $\tilde{\tau}_{\rm
  ref}$ and the radiation pressure $P_{\rm ref}$ at $R_{\rm ref}$ are
known. Following \citet{luc1:93} we assume that $T=T_{\rm ref}$ for
$\tilde{\tau}_{\rm ref}=2/3$. Moreover, we adopt $P_{\rm
  rad}=\frac{4\sigma}{3c}T^4$ for the optically thick regime and
obtain
\begin{equation}
 P_{\rm ref} = \frac{4\sigma}{3c}T_{\rm ref}^4 = \frac{4}{3c} F_{\rm ref},
\end{equation}
and Eq.\,\ref{eq:pradf} becomes
\begin{equation}
\label{eq:peq}
  P_{\rm rad} = P_{\rm ref} \times \left( \frac{1}{2} + \frac{3}{4} \tilde{\tau}\right),
\end{equation}
or
\begin{equation}
\label{eq:teq}
T^4 = T_{\rm ref}^4 \times \left(\frac{1}{2} + \frac{3}{4} \tilde{\tau}\right).
\end{equation}
This result is identical to the more general relation from
\citet{luc1:71,luc1:76,luc1:93}, for $\tilde{\tau} > 2/3$.

\subsection{Numerical tests}
\label{sec:tests}

\begin{figure}[]
  \parbox[b]{0.49\textwidth}{\center{\includegraphics[scale=0.47]{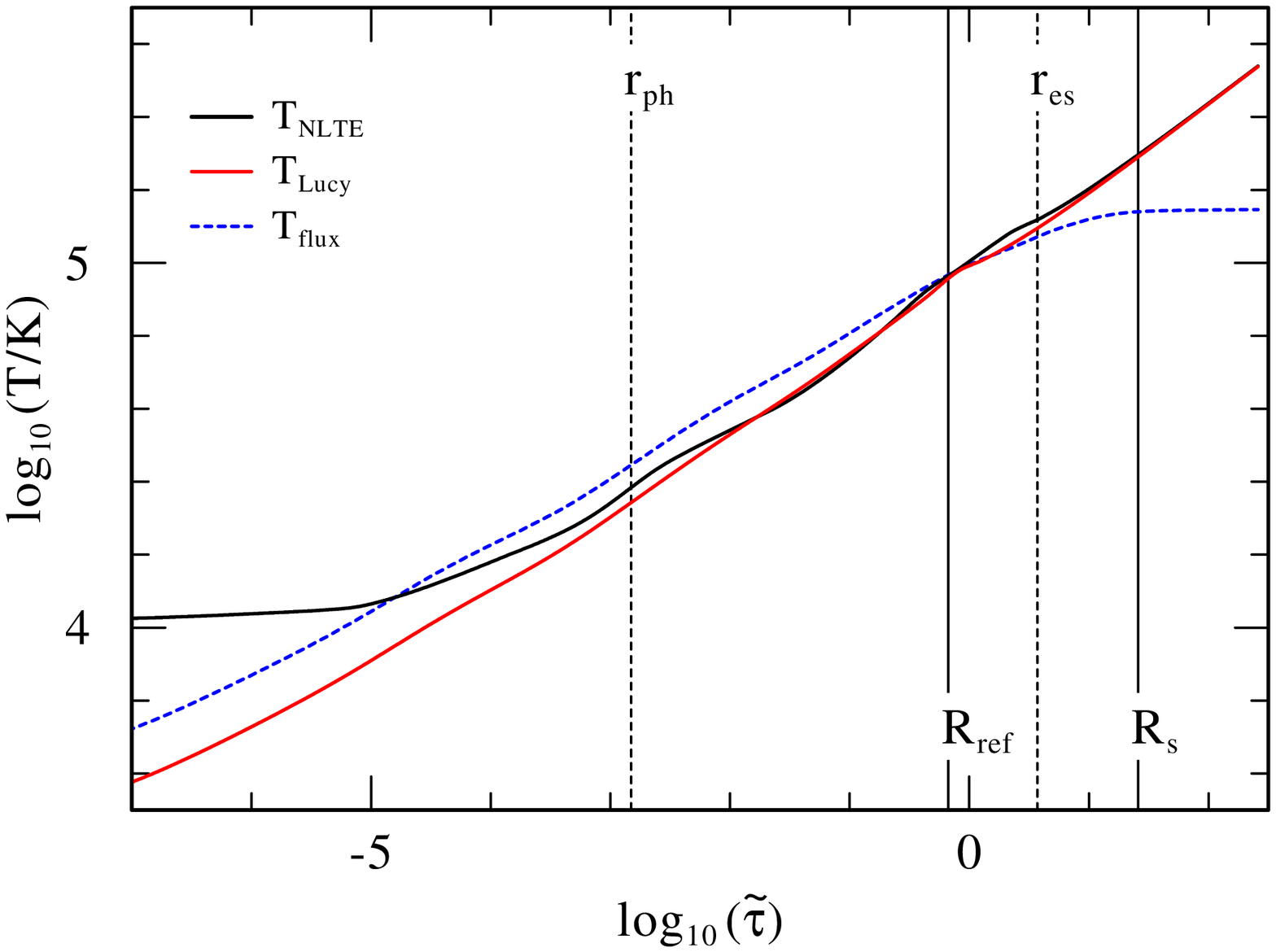}}}
  \caption{Temperature structure $T(\tilde{\tau})$ from the WC star
    model by \citet{gra1:05}. $T_{\rm NLTE}$ (black) is the
    temperature computed in non-LTE within the model, and $T_{\rm
      Lucy}$ (red) is obtained from the relation by \citet{luc1:93}
    which is identical with Eq.\,\ref{eq:teq} for $\tilde{\tau} >
    2/3$.  $T_{\rm flux}$ (blue dashed) denotes the local flux
    temperature. The dashed vertical lines indicate the classical
    photosphere $r_{\rm ph}$ with a flux-mean optical depth
    $\tau=2/3$, and the electron-scattering photosphere $r_{\rm es}$
    with an electron-scattering optical depth $\tau_{\rm es}=2/3$.
    Solid vertical lines indicate the sonic radius $R_{\rm s}$ and the
    reference radius $R_{\rm ref}$ with $\tilde{\tau}=2/3$ where
    $T_{\rm Lucy}=T_{\rm flux}$ is adopted (cf.\ Eq.\,\ref{eq:tref}).}
  \label{fig:temp}
\end{figure}

To verify the applicability of Eqs.\,\ref{eq:taus} and \ref{eq:teq} we
use numerical models for the atmospheres/winds of WR stars by
\citet{gra1:05,gra1:08}. These models compute the detailed radiation
field in the co-moving frame of reference (CMF) using the method by
\citet{koe1:02}, and solve the equations of radiative and statistical
equilibrium in non-LTE to compute the electron temperature $T(r)$ and
the atomic level populations $n_i(r)$ \citep[][]{gra1:02,ham1:03}.
$\rho(r)$ and $\varv(r)$ are obtained simultaneously with the non-LTE
quantities from a precise iterative solution of Eq.\,\ref{eq:motion}
using the radiative acceleration $g_{\rm rad}$ as obtained from an
explicit integration of $\kappa_\nu \times F_\nu$ in
Eq.\,\ref{eq:flux-mean} \citep[cf.][]{gra1:05}.

The numerical models provide $\rho(r)$, $\varv(r)$, $T(r)$, and the
populations $n_i(r)$ for each atomic energy level $i$. The latter are
used to compute opacities $\kappa_\nu(r)$, emissivities $\eta_\nu(r)$,
and the intensity $I_\nu(\mu,r)$ in the CMF. These quantities are
computed on a grid that usually comprises $\sim 10^5$ frequency
points, and $\sim 70$ grid points in radius $r$ and angle $\mu$. The
fluxes $F_\nu$ are then obtained by integration of $I_\nu \times \mu$
over $\mu$. $\kappa_F$ and $g_{\rm rad}$ are evaluated in the CMF
using Eqs.\,\ref{eq:prad}, \ref{eq:flux} and \ref{eq:flux-mean}.

We start with the hydrodynamic WC star model from \citet{gra1:05}.
This model has a very compact core and high $T_\star$.  The inner
boundary is located at a stellar radius $R_\star = 0.905\,R_\odot$
corresponding to $T_\star=140$\,kK for $\log(L/L_\odot)=5.45$.

To estimate $T(\tilde{\tau})$ from Eq.\,\ref{eq:teq} we need to
determine $T_{\rm ref}$, i.e.\ we have to find the reference radius
$R_{\rm ref}$ with $\tilde{\tau}_{\rm ref} = 2/3$. As $\rho(r)$ and
$\kappa_F(r)$ are given within the model, this can be done iteratively
by varying $R_{\rm ref}$ and integrating Eq.\,\ref{eq:taumod} until
\begin{equation}
  \tilde{\tau}_{\rm ref}
 = \int_{R_{\rm ref}}^\infty \rho\kappa_F\frac{R_{\rm ref}^2}{r^2} \dint r
 = \frac{2}{3}.
\end{equation}
For given $R_{\rm ref}$ the flux $F_{\rm ref}$ and flux temperature
$T_{\rm ref}$ can be computed from Eq.\,\ref{eq:tref}.  For our
example model we find $R_{\rm ref}=2.28\,R_\star$ and $T_{\rm
  ref}=92473$\,K. This value compares very well with the non-LTE
temperature $T(R_{\rm ref})=91976$\,K as computed from the condition
of radiative equilibrium within the model (cf.\ Fig.\,\ref{fig:temp}).
This result nicely supports the assumption that $T=T_{\rm ref}$ for
$\tilde{\tau}=2/3$.

In Fig.\,\ref{fig:temp} we compare $T(\tilde{\tau})$ as computed from
Eq.\,\ref{eq:teq} with the non-LTE temperature from the numerical
model.  In particular in deep atmospheric layers ($\tilde{\tau}
\gtrsim 10$) both temperatures agree remarkably well.  E.g., at the
sonic radius $R_{\rm s}$ ($=1.03\,R_\star$) the agreement is better
than 2\%. Large differences only occur in the outermost layers, beyond
the `classical' photospheric radius $r_{\rm ph}$ (defined with respect
to the flux-mean optical depth $\tau(r_{\rm ph}) = 2/3$). In our
example $r_{\rm ph}$ is located at $25.0\,R_\star$.  This is very far
out in the wind, but still within the formation region of the strong
WR emission lines.

The photosphere defined with respect to the electron scattering
opacity $\kappa_{\rm es}$ is located below $R_{\rm ref}$ at $r_{\rm
  es} = 1.41\,R_\star$. $r_{\rm es}$ is of the same order of magnitude
as $R_{\rm ref}$, and can be estimated easily using analytical
relations e.g.\ by \citet{del1:82}. For this reason we will use it in
Sect.\,\ref{sec:rough} to obtain very coarse, but purely empirical
estimates of the sonic point conditions.

\subsection{Representation by $\beta$-law models}
\label{sec:beta}

In the remainder of this work we want to investigate WR stars in
general, i.e.\ without having access to radiation-hydrodynamic models
as in Sect.\,\ref{sec:tests}. To this purpose we {\em assume} that the
winds of our sample stars are radiatively driven, and derive the
radiative acceleration $g_{\rm rad}(r)$ and mean opacity $\kappa_F(r)$
adopting a $\beta$-type velocity distribution
\begin{equation}
\label{eq:betalaw}
  \varv(r) = \varv_\infty\left( 1 - \frac{R_0}{r}\right)^\beta.
\end{equation}
Here $\beta$ is the wind acceleration parameter (usually of the order
of one), and $R_0$ ($\approx R_\star$) the radius parameter. $R_0$ is
adjusted to connect $\varv(r)$ continuously to an exponential velocity
law with $\varv(r) \propto \exp((r-R_\star)/H)$ at the inner boundary
of the model atmosphere. We note that in some cases $R_{\rm s}$ may be
located in the exponential regime, so that the adopted scale height
$H$ may influence our results.

To infer the sonic-point conditions in an analogous way as in
Sect.\,\ref{sec:tests}, we compute $\varv (\ddif\varv/\ddif r)$ from
the given velocity structure $\varv(r)$, and compute the flux-mean
opacity $\kappa_F(r)$ which is consistent with $\varv(r)$ from
Eqs.\,\ref{eq:motion} and \ref{eq:prad}. This is easily possible as
the gas pressure gradient $\dint P_{\rm rad}/\dint r$ in
Eq.\,\ref{eq:motion} is nearly negligible in the supersonic region,
i.e.\ it could be set to zero.  Nevertheless we compute $P_{\rm gas}$
for the temperature resulting from Eq.\,\ref{eq:teq}, and correct for
it in an iterative way.

We apply this method to a model that resembles the properties of the
hydrodynamic model from Sect.\,\ref{sec:tests} but uses a velocity law
with $\beta=1$ \citep[model D from][]{gra1:05}. In particular, the
model has (almost) the same wind efficiency $\eta = 2.5$ as the
hydrodynamic model. The resulting optical-depth scales and sonic-point
conditions for the two models are given at the bottom of
Tab.\,\ref{tab:numeric}.

Although the velocity structure of the hydrodynamic model is
significantly different from a $\beta$-type velocity law \citep[cf.\
Fig.\,8 in][]{gra1:05} we obtain a very similar wind optical depth
$\tau_{\rm s}$($\approx 10$) as for the hydrodynamical model.
Consequently the correction factor $f$ in Eq.\,\ref{eq:taus} is almost
identical for both models, and the sonic-point conditions are almost
the same. For the hydrodynamic model we find $R_{\rm
  s}=1.027\,R_\star$ and $T_{\rm s}=195160$\,K, and for the
$\beta$-law model $R_{\rm s}=1.021\,R_\star$ and $T_{\rm
  s}=201817$\,K. This result is in line with \citet{vin1:12} who found
that the correction factor $f$ mainly depends on the ratio
$\varv_\infty/\varv_{\rm esc}$, and not on the detailed velocity
structure. For our upcoming analysis in Sect.\,\ref{sec:application}
we will thus adopt $\beta$-type velocity distributions.

\section{Application to Galactic WC stars}
\label{sec:application}

Here we apply the methods described in Sect.\,\ref{sec:theory} to a
large sample of WC stars. WC stars are the naked cores of massive
stars that have been stripped off their H-rich layers during their
previous evolution. They show the products of He-burning at their
surface \citep[e.g.][]{gra1:98} and have particularly strong optically
thick winds.  The vast majority of these stars is believed to be in
the phase of core He-burning, with only very few cases in later
burning stages just before a supernova explosion \citep[this class of
objects may be represented by the minority of WO stars;
cf.][]{yoo1:12}. Due to the strong temperature-sensitivity of
He-burning, WC stars are expected to have very similar core
temperatures and will thus follow a well-defined mass-luminosity
relation \citep{lan1:89}.

The complete sample of Galactic WC and WO stars with available optical
spectroscopy has recently been analysed by \citet{san1:12}.  Here we
use the stellar/wind parameters (i.e.\ luminosities $L$, stellar
temperatures $T_\star$, mass-loss rates $\dot{M}$, and terminal wind
velocities $\varv_\infty$) of the 45 putatively single WC/WO stars
from this work, combined with results for Galactic and LMC WC/WO stars
by
\citet{hil1:99,dem1:00,des1:00,sma1:01,cro2:06,cro1:02,cro1:00,gra1:02,gra1:05}
which are based on atmosphere models including Fe line-blanketing.  In
total we have a sample of 61 sets of stellar/wind parameters for WC
stars for which masses can be obtained from the mass-luminosity
relation by \citet{lan1:89}.

For this parameter set we estimate sonic-point densities and
temperatures as described in the previous section.  In
Sect.\,\ref{sec:rough} we start with a very coarse but almost purely
empirical approach to compute the sonic-point conditions. Then we
continue in Sect.\,\ref{sec:numeric} with a more detailed analysis
that is based on numerical models as described in
Sect.\,\ref{sec:beta}.

\subsection{Semi-empirical estimates of the sonic point conditions}
\label{sec:rough}

To obtain the sonic-point temperature $T_{\rm s}=T(R_{\rm s})$ from
Eq.\,\ref{eq:teq} we need to determine $T_{\rm ref}$ and
$\tilde{\tau}_{\rm s} = \tilde{\tau}(R_{\rm s})$. Here we start with a
simplified, but purely empirical method to estimate these quantities.
To this purpose we assume that the reference radius $R_{\rm ref}$ is
of the same order of magnitude as the radius of the
electron-scattering photosphere $r_{\rm es}$ where $\tau_{\rm es}=2/3$
(cf.\,Sect.\,\ref{sec:tests}). $\tau_{\rm es}$ can be computed
analytically. Using the relations by \citet{del1:82} we give
$\tau_{\rm es}$ in Eqs.\,\ref{eq:taues_beta} and \ref{eq:taues1} using
the parameter $C = \dot{M}/(R_\star \varv_\infty)$ which is
proportional to the product of wind density and stellar radius.
Furthermore $\tau_{\rm es}$ is proportional to the electron scattering
opacity $\kappa_{\rm es} = (0.22\,{\rm cm}^2/{\rm g}) \times (1+X)$,
where we assume a fully ionized plasma with hydrogen mass fraction
$X$. For a given value of $\beta\ne 1$ and $F=r_{\rm es}/R_\star$ the
condition $\tau_{\rm es}=2/3$ then reads
\begin{equation}
  \label{eq:taues_beta}
  \tau_{\rm es} = - \kappa_{\rm es} C (1-\beta) \left[\left(1-1/F\right)^{1-\beta} -1\right] = 2/3,
\end{equation}
and for $\beta = 1$
\begin{equation}
  \label{eq:taues1}
  \tau_{\rm es} = - \kappa_{\rm es} C \ln(1-1/F) = 2/3.
\end{equation}

In Tab.\,\ref{tab:rough} we list estimates of wind parameters based on
this approach. They are determined in the following way. $T_{\rm ref}$
is computed from Eq.\,\ref{eq:tref} assuming that $R_{\rm ref} =
r_{\rm es}$ using $F=r_{\rm es}/R_\star$ from Eq.\,\ref{eq:taues1}.
$\tilde{\tau}_{\rm s}$ is approximated by $\tilde{\tau}_{\rm s}
\approx \tau_{\rm s} \times F^2$, and the correction factor
$f$ in Eq.\,\ref{eq:taus} is approximated by
$f\sim\frac{\Gamma-1}{\Gamma}\sim\frac{\varv_\infty/\varv_{\rm
    esc}}{\varv_\infty/\varv_{\rm esc}+1}$ \citep[cf.][]{vin1:12}.
This leads to
\begin{equation}
  \tilde{\tau}_{\rm s} \approx \tau_{\rm s} \times F^2 
= \frac{\eta}{f}\times F^2
\approx \frac{\dot{M}\varv_\infty}{(L/c)} \times \left(1 + \frac{\varv_{\rm esc}}{\varv_\infty}\right) \times F^2.
\end{equation}

\begin{figure}[]
  \parbox[b]{0.49\textwidth}{\center{\includegraphics[scale=0.4]{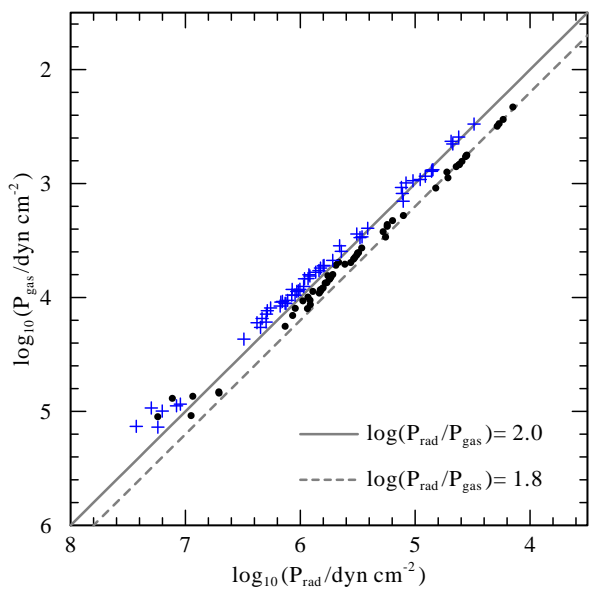}}}
  \caption{Sonic-point conditions for our sample of putatively single
    WC/WO stars in the Galaxy and LMC. Density and temperature at the
    sonic point are expressed in terms of gas and radiation pressure
    $P_{\rm gas}$ and $P_{\rm rad}$. Blue crosses indicate results
    from our empirical analysis in Sect.\,\ref{sec:rough}, black dots
    indicate estimates based on $\beta$-type wind models from
    Sect.\,\ref{sec:numeric}. Both methods indicate a sonic-point
    relation with $P_{\rm rad}/P_{\rm gas} = const.$ {\changedA
      Uncertainties for the semi-empirically determined data points
      are discussed in Sect.\,\ref{sec:uncertainties}. For the ratio
      of $P_{\rm rad}/P_{\rm gas}$ they are likely of the order of
      0.3\,dex, but may be higher for individual positions {\em along}
      the indicated relations.}}
  \label{fig:relation}
\end{figure}

At this point we have expressed $\tilde{\tau}_{\rm s}$ by quantities
that are determined in spectral analyses. Based on $\tilde{\tau}_{\rm
  s}$ we then obtain $T_{\rm s}$ from Eq.\,\ref{eq:teq}, and
$\rho_{\rm s}$ from Eq.\,\ref{eq:cont} assuming that $r_{\rm s}\approx
R_\star$. From $T_{\rm s}$ and $\rho_{\rm s}$ we compute the gas
pressure $P_{\rm gas}$ and radiation pressure $P_{\rm rad}$ at the
sonic point assuming a mean molecular weight $\mu=4/3$. The resulting
values are listed in Tab.\,\ref{tab:rough} and plotted in
Fig.\,\ref{fig:relation}. Please note that we have inverted the axes
in Fig.\,\ref{fig:relation} so that the lower left corner of the plot
corresponds to high densities and temperatures as they are found deep
inside the stellar envelope, while the upper right corner corresponds
to layers further outside.

In Fig.\,\ref{fig:relation} it can be seen that all stars in our
sample follow a relation with $\log(P_{\rm rad}/P_{\rm gas}) \approx
2$, i.e.\ at the sonic point radiation pressure {\em generally}
dominates gas pressure roughly by a factor 100.  The existence of such
a relation is surprising as our sample contains a variety of WC/WO
subtypes with very different wind properties and metallicities. This
relation may thus be a ubiquitous property of optically thick winds.

The temperature and density estimates obtained in the present section
are approximative but based on universal relations. The only place
where the detailed velocity structure goes in (in the form of the
parameter $\beta$) is the computation of $R_{\rm ref}$ based on
Eq.\,\ref{eq:taues1}. A main source of uncertainty may be the
assumption that $R_{\rm ref}=r_{\rm es}$. In our example in
Sect.\,\ref{sec:tests} we have seen that $R_{\rm ref}>r_{\rm es}$.
The reference temperatures $T_{\rm ref}$, and consequently also the
sonic point temperatures $T_{\rm s}$ obtained in this section are thus
likely too high.

\subsection{Numerical estimates}
\label{sec:numeric}

In this section we use the numerical modelling approach described in
Sect.\,\ref{sec:beta} to estimate the sonic-point conditions for our
sample stars. We adopt a $\beta$-type velocity law
(Eq.\,\ref{eq:betalaw}) with $\beta=1$ for $\varv(r)$ and compute all
optical depth scales using the opacity $\kappa_F(r)$ that is
consistent with the corresponding wind acceleration $\varv(\ddif\varv
/ \ddif r)$, i.e.\ we assume that the winds are radiatively driven.

The results are listed in Tab.\,\ref{tab:numeric} and plotted in
Fig.\,\ref{fig:relation} in the same form as the results from
Sect.\,\ref{sec:rough}. With the numerical method the relation from
Sect.\,\ref{sec:rough} is retained, but the obtained temperatures are
lower because the reference radii $R_{\rm ref}$ are larger than
$r_{\rm es}$. Consequently the relation for $P_{\rm rad}/P_{\rm gas}$ shifts by
$\sim$\,0.2\,dex with respect to the data points from
Sect.\,\ref{sec:rough}. In view of the large differences between both
approaches this demonstrates the robustness of this result.

As we will discuss later, the relation for $P_{\rm rad}/P_{\rm gas}$
obtained in this way imposes a boundary condition on the stellar
envelope at the sonic point. More precisely, it reflects a boundary
condition that is imposed by the presence of an optically thick,
radiatively driven wind. According to our present semi-empirical
results this `wind condition' has the form
\begin{equation}
\label{eq:sonic}
\frac{P_{\rm rad}}{P_{\rm gas}} \approx 80.
\end{equation}

\section{An artificial model sequence for WC/WO stars}
\label{sec:sequence}

In order to investigate the general applicability of our results, as
well as their sensitivity to uncertainties in the observed stellar
parameters, we construct a model sequence for WC/WO stars that is
suitable for a systematic analysis. To this purpose we introduce the
concept of the transformed mass-loss rate $\dot{M}_{\rm t}$ in
Sect.\,\ref{sec:mtrans}. In Sect.\,\ref{sec:obs} we discuss the
observed properties of WC/WO stars and define an artificial WC/WO
sequence.  In Sect.\,\ref{sec:uncertainties} we vary input parameters
and model assumptions to investigate the reliability of our method.

\subsection{The transformed mass-loss rate}
\label{sec:mtrans}

A remarkable property of massive WC stars is their homogeneous
spectroscopic appearance. For a given spectral subtype (i.e.\ for
given $T_\star$) their normalized spectra are almost invariant
regardless of their luminosity. The stellar/wind parameters of WC
stars thus follow a scaling relation which preserves the equivalent
widths of their emission lines for given $T_\star$. If the luminosity
$L$ (or equivalently the radius $R_\star$) are changed, this scaling
relation has to ensure that the wind parameters (mass-loss rate
$\dot{M}$, terminal wind speed $\varv_\infty$ and clumping factor $D$)
are adapted in a way that the line equivalent widths stay constant.

Based on numerical results, such a scaling relation has been
introduced by \citet[][]{sch1:89} via the so-called transformed radius
$R_{\rm t}$. This parameter is of large benefit for the analysis of
emission-line stars as it reduces the number of free parameters by
two, and enables a two-step analysis where the normalized spectrum is
modeled in step one, and the flux-distribution (including interstellar
extinction) in step two.

\citet{ham1:98} explained this invariance with the dominance of
recombination processes for the line emission in WR stars. As
recombination is an $n^2$ process, the line emissivity $j$ per unit
volume scales with $j\propto n^2$, and its spatial mean goes with
$\bar{j}\propto n^2/D$ if clumping is taken into account (cf.
Sect.\,\ref{sec:clumping}). The observed line flux $F_l$ is given by
the product of the mean emissivity and the line-emitting volume, i.e.\
$F_l \propto \bar{j} \times V$. At this point \citet{ham1:98} assumed
that $V\propto R_\star^3$, i.e.\ the radial size $\Delta r$ of the
line-emitting region in the spherically symmetric wind scales as
$\Delta r = \Delta r' \times R_\star$ with $\Delta r'=const.$

Here we show that this assumption is indeed correct if $\Delta r$ is
determined by photoionization equilibrium. In this case the ionization
rate is balanced by the recombination rate, i.e.\ $F \times n_s
\propto n^2$, where $F$ denotes the ionizing flux, $n_s$ the density
of the subordinate ionization stage, and $n$ the density of the main
ionization stage (which is almost equal to the total particle
density). If we adopt a radial scaling with $r = r' \times R_\star$,
then $F(r')$ is invariant if also the optical depth $\tau(r') \propto
n_s(r')\times \Delta r /D$ in the ionizing continuum is preserved.
With the scaling relations for $\Delta r$ and $n_s$ from above this
relation becomes $\tau(r') \propto n_s(r')\times \Delta r /D \propto
n^2(r') \times R_\star /D =const$. To preserve the optical depth scale
$\tau(r')$ we thus need to fulfill the scaling relation $n^2(r')
\times R_\star /D =const.$

Notably, in this case we have $F_l \propto \bar{j} \times V \propto
n^2(r') \times R_\star^3/D \propto R_\star^2$. The line flux
$F_l\propto R_\star^2$ thus scales in the same way with the stellar
radius as the luminosity $L\propto R_\star^2$, i.e.\ line equivalent
widths are preserved. To express this scaling relation in terms of
standard wind parameters we use Eqs.\,\ref{eq:cont} and
\ref{eq:betalaw}, i.e.\ $\dot{M} \propto n(r') \varv(r') r^2/D \propto
n(r') \varv_\infty R_\star^2/D$. This finally leads to $n^2(r') \times
R_\star /D \propto (\dot{M}/\varv_\infty)^2 R_\star^3 D=const.$ In the
following we express this scaling relation via the transformed
mass-loss rate
\begin{equation}
  \label{eq:mtrans}
  \dot{M}_{\rm t} = {\dot{M} \sqrt{D}} \times
{\left(\frac{1000\,{\rm km\,s}^{-1}}{\varv_\infty}\right) \left(\frac{10^6 L_\odot}{L}\right)^{3/4}}.
\end{equation}
For a star with given emission line equivalent widths $\dot{M}_{\rm
  t}$ denotes the mass-loss rate that the star {\em would} have if it
had $L=10^6 L_\odot$, $\varv_\infty=1000\,{\rm km\,s}$, and $D=1$.  We
note that this definition is fully equivalent to the relation
resulting from the transformed radius $R_{\rm t}$ as introduced by
\citet[][]{sch1:89,ham1:98}. In the context of our present work
$\dot{M}_{\rm t}$ is, however, more meaningful as it is directly related
to the stellar mass-loss rate.

\subsection{Stellar/wind parameters of WC/WO stars}
\label{sec:obs}

\begin{figure}[]
  \parbox[b]{0.49\textwidth}{\center{\includegraphics[scale=0.42]{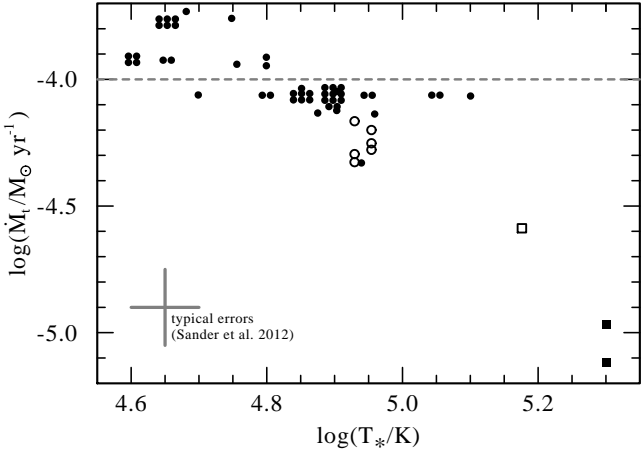}}}
  \caption{Transformed mass-loss rates $\dot{M}_{\rm t}$
    (Eq.\,\ref{eq:mtrans}) versus stellar temperatures $T_\star$ for
    our sample of putatively single WC (circles) and WO stars
    (squares). Filled symbols indicate objects in the Galaxy, empty
    symbols objects in the LMC. Some symbols have been shifted to
    avoid overlaps. The grey dashed line indicates the transformed
    mass-loss rate adopted for our artificial model sequence.}
  \label{fig:mtrans}
\end{figure}

\begin{figure}[]
  \parbox[b]{0.49\textwidth}{\center{\includegraphics[scale=0.42]{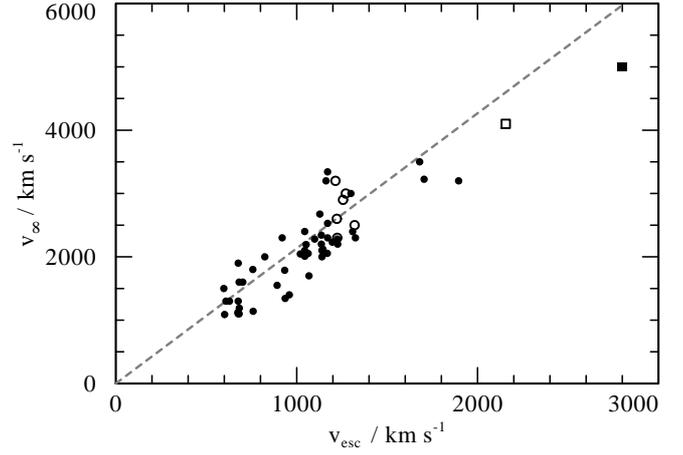}}}
  \caption{Terminal wind velocities $\varv_\infty$ vs.\ escape
    velocities $\varv_{\rm esc}$ for our sample of WC/WO stars.
    Symbols are the same as in Fig.\,\ref{fig:mtrans}. {\changedA
      Typical uncertainties are of the order $\Delta \varv_\infty =
      \pm 10\%$ and $\Delta \varv_{\rm esc} = \pm 15\%$.} The grey
    dashed line indicates the relation with $\varv_\infty/\varv_{\rm
      esc}=1.6$ as adopted for our artificial model sequence.}
  \label{fig:vinf}
\end{figure}

In Fig.\,\ref{fig:mtrans} we plot $\dot{M}_{\rm t}$ vs.\ $T_\star$ for
the stars in our sample. Notably, the Galactic WC stars are lying in a
regime with $\log(\dot{M}_{\rm t}) \approx -4$ (in $\msunpyr$), with a
tendency of slightly higher mass-loss rates for late subtypes.  This
relation is equivalent to the relation for the transformed radius
$R_{\rm t}$ with $R_{\rm t}\propto T_\star^{-2}$ from
\citet{bar2:06,san1:12}. Only the extremely hot Galactic WO stars form
a separate group with much lower $\log(\dot{M}_{\rm t}) \approx -5$.

The LMC WC stars have slightly lower $\dot{M}_{\rm t}$ than the
Galactic ones, and the only LMC WO star is located somewhat between
the Galactic WC and WO stars.

Based on the few WO stars and the slight excess for late subtypes,
Fig.\,\ref{fig:mtrans} may even suggest a steep decrease of
$\dot{M}_{\rm t}$ with increasing $T_\star$. It is, however, not clear
whether this decrease is related to a metallicity dependence of WR
mass-loss \citep[as theoretically predicted by][]{vin1:05,gra1:08}, or
if it is an intrinsic property of WC/WO stars.  Observationally, late
WC subtypes are mainly found in high-metallicity environments, while
early subtypes dominate at low metallicities as in the LMC. In any
case most Galactic WC stars comply with $\log(\dot{M}_{\rm t}) \approx
-4$ as indicated by the dashed grey line in Fig.\,\ref{fig:mtrans}. In
the following we adopt this value for our artificial model sequence.

In Fig.\,\ref{fig:vinf} we plot $\varv_\infty$ vs. $\varv_{\rm esc}$
for our sample stars. While $\varv_\infty$ can be inferred directly
from the observed blue edges of P-Cygni absorption troughs,
$\varv_{\rm esc}$ is determined from the stellar luminosity $L$ and
temperature $T_\star$ as obtained from spectral analyses. Based on
these two quantities we compute the radii $R_\star$. The stellar
masses $M$ follow from $L$, and to a lesser extent from the
obtained/adopted surface abundances, using the relation from
\citet{lan1:89}. The resulting escape velocities $\varv_{\rm esc} =
\sqrt{2GM/R_\star}$ are listed together with the other stellar
parameters in Tab.\,\ref{tab:numeric}.  Fig.\,\ref{fig:vinf} clearly
suggests a relation between $\varv_\infty$ and $\varv_{\rm
  esc}$\footnote{\changedA Observational uncertainties for
  $\varv_\infty$ are likely of the order of $\pm 10\%$. For
  $\varv_{\rm esc}$ we obtain with $L\propto M^{\gamma}$ and
  $R_\star^2 \propto L/T_\star^4$ that $\varv_{\rm esc} \propto
  \sqrt{M/R_\star} \propto L^{1/(2\gamma)-1/4}/T_\star$. For the
  present sample we estimate $\gamma \sim 1.8$ so that $\varv_{\rm
    esc} \propto L^{0.03}/T_\star$. With typical errors of $\Delta
  \log(L)=\pm 0.3$ and $\Delta \log(T_\star)=\pm 0.05$ from
  \citet{san1:12} we obtain an uncertainty of $\Delta \varv_{\rm esc}
  = \pm 15\%$.}.  Again, this picture may be complicated by a possible
dependence on metallicity \citep[cf.][]{gra1:08}, and the low numbers
of extremely compact WO stars. For our artificial model sequence we
adopt $\varv_\infty/\varv_{\rm esc}\sim1.6$ as indicated by the dashed
grey line in Fig.\,\ref{fig:vinf}.

We note that the existence of a relation with $\varv_\infty/\varv_{\rm
  esc}=const.$ is plausible as it implies that the mechanical wind
energy is of a similar order of magnitude as the gravitational wind
energy (cf.\,Sect.\,\ref{sec:tiring}). The observed relation in
Fig.\,\ref{fig:vinf} thus supports the existence of a broad range of
gravitational energies for WC/WO stars. This means that the stellar
radii also span a broad range, in accordance with the $R_\star$
obtained from spectral analyses. In contrast, theoretically predicted
values \citep[e.g.\ by][]{lan1:89} span a much narrower range. The
observed relation in Fig.\,\ref{fig:vinf} thus supports our scenario
2) from Sect.\,\ref{sec:radii} where the spectroscopically determined
radii reflect the actual surface radii, and the sub-surface layers of
many WC stars are substantially inflated.

\subsection{Dependence on stellar/wind parameters}
\label{sec:uncertainties}

As the obtained sonic-point relation (Eq.\,\ref{eq:sonic}) depends on
stellar parameters as input, we investigate here its sensitivity to
uncertainties in these parameters.

Apart from $\varv_\infty$ which can be reliably determined from
observed UV line profiles, most of our `input' stellar parameters
($\dot{M}$, $L$, and $T_\star$) rely on numerical modelling and may
suffer from systematic uncertainties.  While uncertainties in the
empirical mass-loss rates $\dot{M}$ are probably dominated by the
effects of clumping and porosity (cf.\ Sect.\,\ref{sec:clumping}), $L$
and $T_\star$ may suffer from uncertainties in the model physics,
mainly due to the (in)completeness of the included atomic data. In the
past, Fe-group line blanketing turned out to affect the overall flux
distribution and temperature of WR model atmospheres considerably.
Since its inclusion in modern non-LTE codes \citep{hil1:98,gra1:02}
quantitative results, however, seem to converge.

To investigate the sensitivity of our results to uncertainties and/or
changes in stellar parameters, we construct an artificial model
sequence with the properties discussed in Sect.\,\ref{sec:obs}, i.e.\
with $\log(\dot{M}_{\rm t}/(\msunpyr)) = -4$ and
$\varv_\infty/\varv_{\rm esc} = 1.6$. We cover a temperature range of
$\log(T_\star/{\rm K}) = 4.5\,...\,5.3$ in steps of 0.1\,dex and use a
typical luminosity for WC stars of $\log(L/L_\sun)=5.5$
\citep[e.g.][]{san1:12}.  We note that wind clumping has no direct
effect on our modelling approach, but it affects the mass-loss rates
$\dot{M}$ that follow from Eq.\,\ref{eq:mtrans}. Here we adopt a
clumping factor of $D=10$. For the velocity distribution we adopt a
$\beta$-law with $\beta=1$ which is smoothly connected to an
exponential law with a fixed density scale height $H$ in the
hydrostatic layers (cf.\ Sect.\,\ref{sec:beta}). $H$ is computed for
$r=R_\star$, $T=T_\star$, and $M_{\rm eff}=M (1-\Gamma)$ with
$\Gamma=0.9$. In the following we investigate the effects of changes
in $\dot{M}$, $L$, $\varv_\infty$, $\beta$, and $H$.

\begin{figure}[]
  \parbox[b]{0.49\textwidth}{\center{\includegraphics[scale=0.44]{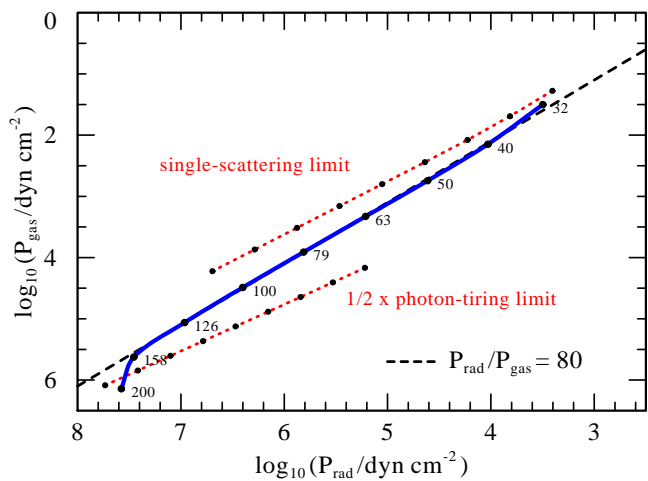}}}
  \caption{Sonic-point conditions for our artificial model sequence
    for WC/WO stars. The blue line indicates the sequence of reference
    models (also denoted as `reference sequence').  Labels indicate
    stellar temperatures $T_\star$ along the reference sequence. The
    black dashed line indicates the wind condition according to
    Eq.\,\ref{eq:sonic}. Red dotted lines indicate relevant mass-loss
    limits (see text).}
  \label{fig:testlimit}
\end{figure}

In Fig.\,\ref{fig:testlimit} we show the inferred sonic-point
conditions for our reference models. First of all, the reference
models form a sequence that reaches from low $P_{\rm rad}$ for the
coolest models to high $P_{\rm rad}$ for the hottest models. In the
following we denote this sequence as the `reference sequence'.
Notably, 8 of the 9 models from the reference sequence lie almost
precisely on a straight line with $P_{\rm rad}/P_{\rm gas}=80$, in
agreement with the wind condition Eq.\,\ref{eq:sonic}.  Furthermore,
Fig.\,\ref{fig:testlimit} shows two model sequences for which the
mass-loss rates have been adjusted to match relevant mass-loss limits,
namely the single-scattering limit
($\eta=\dot{M}\varv_\infty/(L/c)=1$) and 50\% of the photon-tiring
limit ($L_{\rm wind}=L/2$).

As $\eta \approx \tau_{\rm s}$ according to Eq.\,\ref{eq:taus}, the
single-scattering limit marks the mass-loss rate for which the wind
becomes optically thin, i.e.\ 1) the star does not appear as a WR star
anymore and 2) our method is not applicable. Because the $\tau_{\rm
  s}$ in these test models are smaller than for our reference models,
their sonic-point temperatures $T_{\rm s}$ are smaller and the
sequence shifts towards lower values of $P_{\rm rad}$.

The photon-tiring limit marks an upper limit for the mass-loss rates
of stationary radiatively driven winds
\citep[e.g.][]{owo1:04,mar2:09}. It is reached when the (mechanical +
gravitational) wind luminosity equals the radiative luminosity of the
star, i.e.\ $L_{\rm wind} = \dot{M} \times (\varv_\infty^2/2 +
MG/R_\star) = L$. In this case the radiative luminosity present at
$R_\star$ would be completely transformed into wind energy and the
star would appear dark for the observer. At this point our models
would fail. In Fig.\,\ref{fig:testlimit} we thus indicate 50\% of the
photon-tiring limit, in which case $L_{\rm wind} = L/2$, i.e.\ the
radiative luminosity is significantly reduced within the model.
As the temperature is sustained by radiative heating it is
substantially reduced in this case, and falls below the value expected
from Eq.\,\ref{eq:sonic}.  This is the reason why our hottest
reference model deviates from the standard wind condition in
Fig.\,\ref{fig:testlimit}.

\begin{figure}[]
  \parbox[b]{0.49\textwidth}{\center{\includegraphics[scale=0.44]{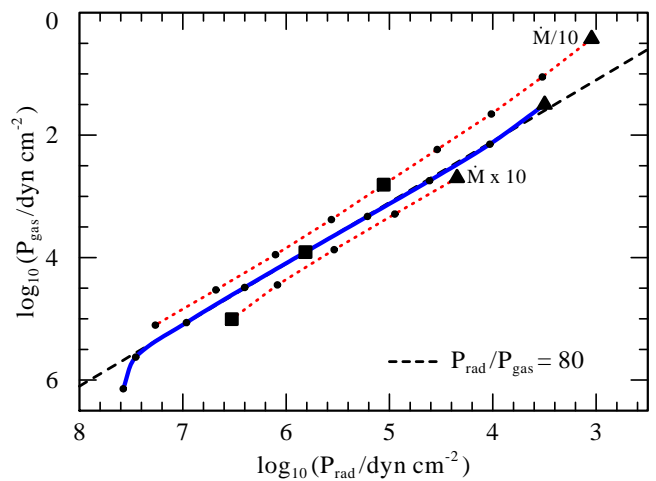}}}
  \caption{Influence of $\dot{M}$ on the sonic-point conditions.  The
    reference sequence (solid blue line) is compared to two test
    sequences for which the mass-loss rates have been varied by a
    factor 10 ({\changedA red dotted} lines). Although the models span
    a factor of 100 in $\dot{M}$ the changes with respect to the wind
    condition Eq.\,\ref{eq:sonic} (black dashed line) are moderate. To
    illustrate the shift in $T_\star$, wind models with
    $T_\star=79$\,kK are indicated by squares and wind models with
    $T_\star=32$\,kK by triangles.}
  \label{fig:testmdot}
\end{figure}

\begin{figure}[]
  \parbox[b]{0.49\textwidth}{\center{\includegraphics[scale=0.44]{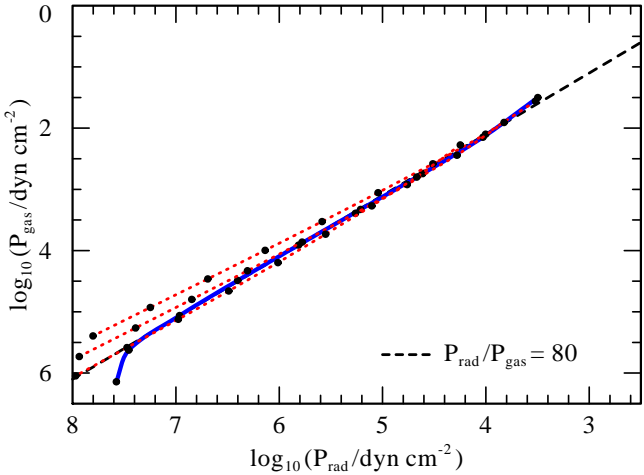}}}
  \caption{Influence of $\varv_\infty$ on the sonic-point conditions.
    The reference sequence (solid blue line) is compared to test
    sequences with $\varv_\infty = 1000$\,km/s, 2000\,km/s, and
    4000\,km/s ({\changedA red dotted} lines).}
  \label{fig:testv}
\end{figure}

\begin{figure}[]
  \parbox[b]{0.49\textwidth}{\center{\includegraphics[scale=0.44]{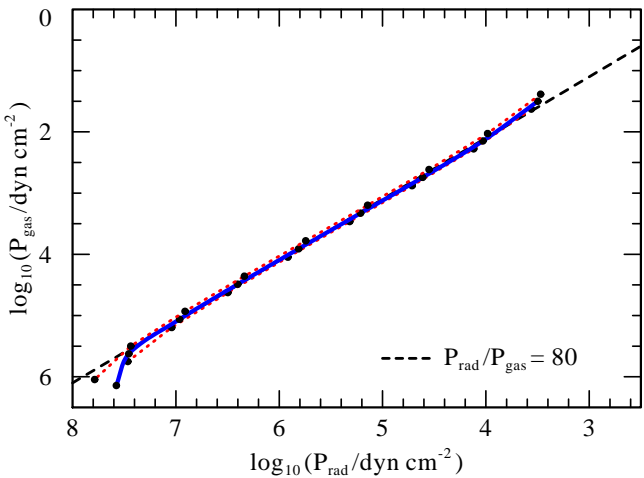}}}
  \caption{Influence of the adopted luminosity $L$ on the sonic-point
    conditions.  The reference sequence (solid blue line) is compared
    to test sequences with $\log(L/L_\odot)=5$ and $\log(L/L_\odot)=6$
    ({\changedA red dotted} lines).}
  \label{fig:testl}
\end{figure}

\begin{figure}[]
  \parbox[b]{0.49\textwidth}{\center{\includegraphics[scale=0.44]{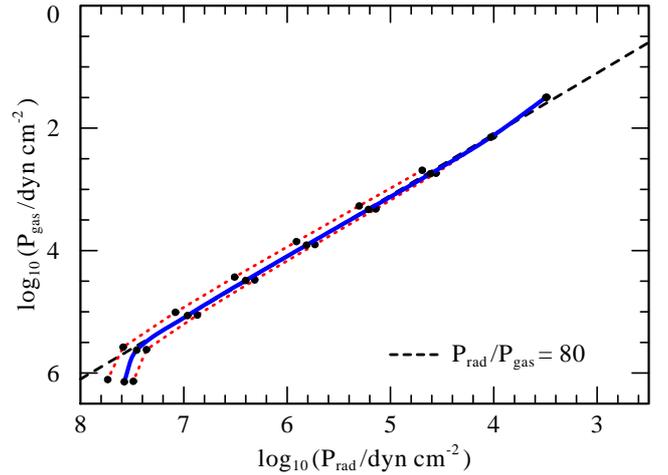}}}
  \caption{Influence of the adopted acceleration parameter $\beta$
    (Eq.\,\ref{eq:betalaw}) on the sonic-point conditions.  The
    reference sequence (solid blue line) is compared to test sequences
    with $\beta=0.5$ and $\beta=2$ ({\changedA red dotted} lines).}
  \label{fig:testbeta}
\end{figure}

\begin{figure}[]
  \parbox[b]{0.49\textwidth}{\center{\includegraphics[scale=0.44]{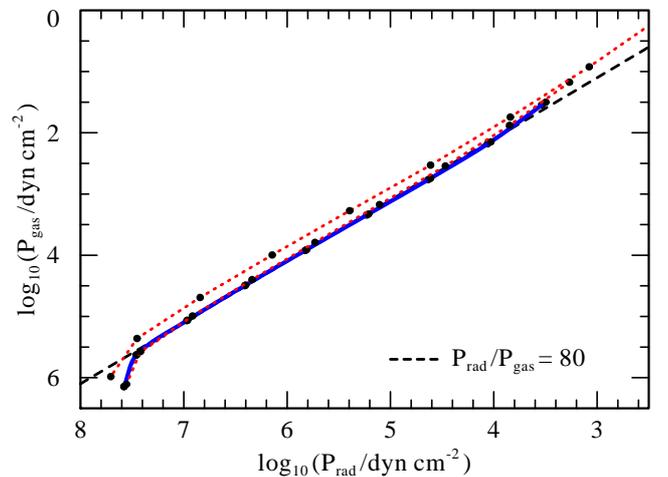}}}
  \caption{Influence of the density scale height $H$ on the
    sonic-point conditions. Here we have varied the adopted Eddington
    parameter $\Gamma$ which has been used to compute $H$.  The
    reference sequence with $\Gamma=0.9$ (solid blue line) is compared
    to test sequences with $\Gamma=0.5$, 0.99, and 0.999 ({\changedA
      red dotted} lines). A significant effect only occurs for
    $\Gamma=0.999$.}
  \label{fig:testh}
\end{figure}

In Figs.\,\ref{fig:testmdot}--\ref{fig:testh} we demonstrate the
effects of the previously discussed parameter changes on the
sonic-point conditions. The strongest effects are caused by changes in
$\dot{M}$ and $\varv_\infty$, which is plausible as they directly
affect $\tau_{\rm s}$ via Eq.\,\ref{eq:taus}.

In Fig.\,\ref{fig:testmdot} we compare the reference sequence with two
test sequences for which $\dot{M}$ (and thus also $\dot{M}_{\rm t}$)
has been changed by a factor 10. All other parameters are kept fixed.
It is remarkable that individual models are predominantly shifted {\em
  along} the sonic-point relation. Models with different mass-loss
rates thus experience strong changes in $T_{\rm s}$ but only a
moderate shift with respect to Eq.\,\ref{eq:sonic} ($\sim$0.3\,dex).
We note that the test performed here is limited, as the 4 hottest
models for high $\dot{M}$ are in conflict with the photon-tiring
limit, and the 4 coolest models for low $\dot{M}$ are below the
single-scattering limit.

In Fig.\,\ref{fig:testv} we vary the velocity structure by changing
$\varv_\infty$, keeping $\dot{M}_{\rm t}$ (and thus the wind density)
and all other other parameters fixed. We compare three test sequences
with $\varv_\infty = 1000$\,km/s, 2000\,km/s, and 4000\,km/s to the
reference sequence. Only for hot models the test sequences do deviate
significantly from the reference sequence.  The models with
$\varv_\infty=1000$\,km/s deviate most. In reality such low values of
$\varv_\infty$ are not observed for early-type WC or WO stars, so that
the expected deviations from Eq.\,\ref{eq:sonic} will be small.

The tests for changes in $L$, $\beta$, and $H$ are shown in
Figs.\,\ref{fig:testl}, \ref{fig:testbeta}, and \ref{fig:testh}. The
effects on the sonic-point conditions turn out to be very small in all
three cases. In Fig.\,\ref{fig:testl} $\log(L/L_\odot)$ has been
varied between 5.0 and 6.0, corresponding to the observed range of
WC/WO luminosities in the Galaxy and LMC. For both luminosities
$\dot{M}$ has been adapted to keep $\dot{M}_{\rm t}$ constant. In
Fig.\,\ref{fig:testbeta} $\beta$ has been varied between 0.5 and 2.0,
and in Fig.\,\ref{fig:testh} the Eddington factor $\Gamma$ that is
used to compute the density scale height $H$ has been varied between
0.5 and 0.999. Only for the most extreme case of $\Gamma=0.999$ is a
significant effect visible, i.e.\ when $H$ is artificially increased
by a factor $10^4$.

{\changedA The observational uncertainties in our sample from
  Sect.\,\ref{sec:application} are dominated by the poorly known
  distances in the Galaxy resulting in an uncertainty of $\Delta
  \log(L)=\pm 0.3$.  Here we have shown that our result for $P_{\rm
    rad}/P_{\rm gas}$ is particularly insensitive to variations $L$.
  The largest uncertainties in $P_{\rm rad}/P_{\rm gas}$ are likely
  due to uncertainties in $\dot{M}$ due to wind clumping. For WR stars
  these may be of the order of 2--3, i.e.\ still only a fraction of
  the factor 100 that we have investigated in
  Fig.\,\ref{fig:testmdot}.  The resulting uncertainty in $P_{\rm
    rad}/P_{\rm gas}$ will be of the order of 0.15\,dex.  The overall
  uncertainty in $P_{\rm rad}/P_{\rm gas}$ e.g.\ for the data points in
  Fig.\,\ref{fig:relation} will probably not exceed $0.3$\,dex.}

We conclude that the wind condition Eq.\,\ref{eq:sonic} is insensitive
to parameter variations within a realistic range. This means that the
existence of an optically thick wind confines the sonic-point
conditions to a narrow strip in the $P_{\rm rad}$--$P_{\rm gas}$
plane, in agreement with our semi-empirical results from
Sect.\,\ref{sec:application}.  Moreover, we find that the
photon-tiring limit puts constraints on the earliest WC/WO subtypes.

\section{Discussion}
\label{sec:discussion}

Here we discuss the relevance of our results for the winds and the
sub-surface structure of stars near the Eddington limit. We start with
a discussion of the photon-tiring effect in Sect.\,\ref{sec:tiring},
as it seems to set constraints on the winds of compact WR stars.  In
Sect.\,\ref{sec:example} we discuss in detail how the wind condition
Eq.\,\ref{eq:sonic} constrains the winds and sub-surface layers near
the Eddington limit.

\subsection{Relevance of the photon-tiring limit for WR stars}
\label{sec:tiring}

{\changedA The photon-tiring limit is reached when the (mechanical plus
  gravitational) wind luminosity becomes equal to the radiative
  luminosity of the star, i.e.\ all photons are used up to drive the
  wind. The photon-tiring effect is included in quantitative
  atmosphere/wind models, and has been identified to play a role for
  the strong winds of WC stars. For quantitative WC models with
  un-clumped winds \citet{gra1:98} found that around 30\% of the
  stellar luminosity are lost to the wind.  Since the advent of wind
  clumping the empirical mass-loss rates have been reduced so that now
  only $\sim$\,10\% of the stellar luminosity go into the wind
  \citep[e.g.][]{cro1:02}.}

For our artificial model sequence in Sect.\,\ref{sec:sequence} we
found that the most compact WR stars are affected by the photon-tiring
effect. The reason is that the wind luminosity depends on the
gravitational well that has to be overcome by the stellar wind. The
mechanical wind luminosity $L_{\rm mech}=(1/2)\dot{M}\varv_\infty^2$
will thus be of the same order of magnitude as the gravitational wind
luminosity $L_{\rm grav}=\dot{M}MG/R_\star=(1/2)\dot{M}\varv_{\rm
  esc}^2$. It is thus plausible to assume that
$\varv_\infty/\varv_{\rm esc}\approx const.$, as we did for our model
sequence in Sect.\,\ref{sec:sequence}.

For the optically thin winds of OB stars relations with
$\varv_\infty/\varv_{\rm esc}=const.$ are theoretically predicted by
\citet{cas1:75}, with a constant ratio that depends on the atomic
properties of the dominating ions in the wind. This is observationally
supported e.g.\ by \citet{lam1:95} who found ratios of 1.3 and 2.6 for
different temperature regimes in the OB star range. For WR stars
\citet{nie1:02} found an increase in $\varv_\infty$ for earlier WR
subtypes, which is in line with our adopted relation in
Sect.\,\ref{sec:sequence}.  In contrast to this \citet{nug1:00} found
$\varv_\infty/\varv_{\rm esc} \sim 1$ for H-poor WN stars, and varying
$\varv_\infty/\varv_{\rm esc} \lesssim 1$ for WC stars, with strongly
deviating values for WO stars. \citeauthor{nug1:00}, however, employed
the small WR radii as predicted from classical stellar structure
models, i.e.\ they did not consider the possibility of an envelope
inflation. Our present results and the ones from \citet{nie1:02} are
based on observed temperatures and radii, and seem to give a more
consistent picture over the complete WC/WO regime.

With our assumption that $\varv_\infty/\varv_{\rm esc}=1.6$ we obtain
$L_{\rm wind}=3.54 \times \dot{M}\varv_{\rm esc}^2/2$. For this case
we find that the hottest models in our sequence are affected by the
photon tiring effect. The observed stars in this regime are all
members of the WO subclass and have $\sim$\,$10\times$ lower
transformed mass-loss rates $\dot{M}_{\rm t}$ than the ones adopted in
our models. The underlying reason for {\changedA the observed mass-loss
  reduction may thus be related to the photon-tiring effect.}

{\changedA As the formation of WR-type winds is not yet fully
  understood it is not clear how such a mechanism would work in
  detail. Ways to reduce the photon tiring effect are, however, 1) a
  reduction of $\dot{M}$, 2) a reduction of $\varv_\infty$, and 3) an
  increase of $R_\star$ through envelope inflation. Following our
  discussion above, a reduction of $\varv_\infty$ may be difficult to
  realize in nature as it would require that $L_{\rm mech}\ll L_{\rm
    grav}$.}  Compact WC stars may thus be in a situation where
{\changedA photon-tiring} forces them to either reduce their mass-loss
rates, as observed for the WO subclass, or to increase their radii so
that a high mass-loss rate can be maintained. In this case they would
appear as normal WC subtypes.

\subsection{Relevance for the theory of optically thick winds}
\label{sec:example}

\begin{figure*}[t!]
  \parbox[b]{0.99\textwidth}{\center{\includegraphics[scale=0.5]{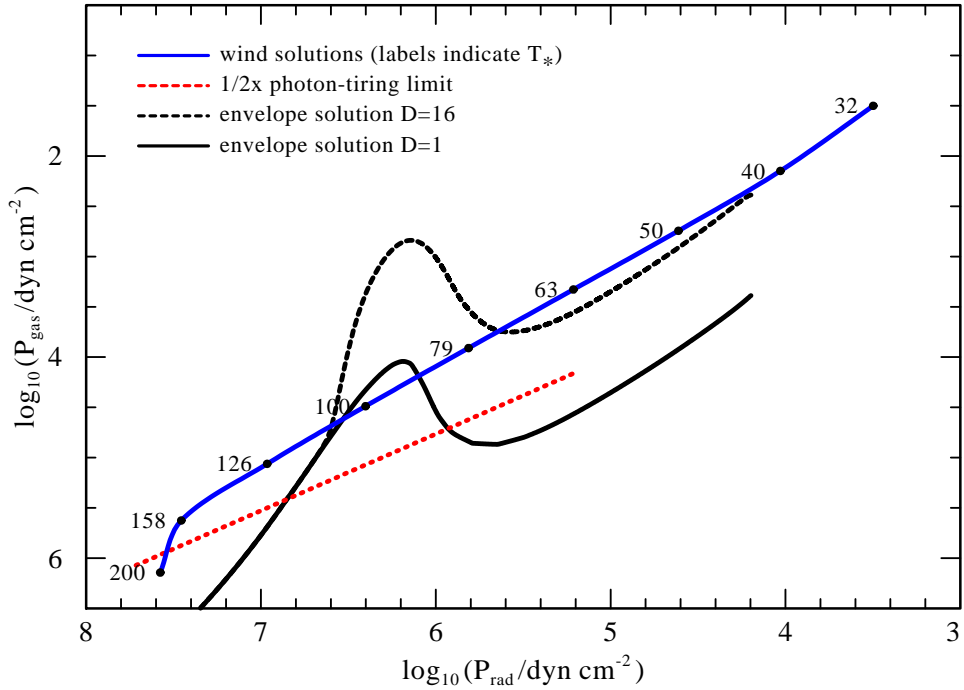}}}
  \caption{Conditions at the sonic point for a $14\,M_\odot$ helium
    star with Galactic metallicity. The solid and dashed black lines
    indicate stellar envelope solutions with different clumping
    factors $D=1$ and 16 in the sub-surface layers. The conditions
    imposed by the optically thick stellar wind are indicated in blue.
    Labels indicate the stellar temperature $T_\star$ of the wind
    solutions. The red dotted line indicates 50\% of the photon-tiring
    limit, i.e.\ the region where half of the stellar luminosity $L$
    is consumed to drive the stellar wind.}
  \label{fig:gamma}
\end{figure*}

The theory of optically thick winds is mainly based on critical point
analyses \citep{pis1:95,nug1:02}. It is assumed that in the deep
layers of optically thick winds the flux-mean opacity $\kappa_F$
equals the Rosseland mean opacity $\kappa_{\rm Ross}(\rho,T)$, so that
the radiative acceleration is given as a function of density and
temperature $g_{\rm rad}(\rho,T)=\kappa_{\rm Ross}(\rho,T) F_{\rm
  rad}/c$ (cf.\ Eq.\,\ref{eq:prad}). It can be shown that in this case
the Eddington limit has to be crossed almost precisely at the sonic
point, i.e.\ the opacity at $R_{\rm s}$ has to match the value of the
Eddington opacity $\kappa_{\rm Edd} = 4\pi c G M / L$ \citep[cf.\
Sect.\ 4.1 in][]{nug1:02}. Furthermore, as the wind has do be
accelerated outward beyond the sonic point, the opacity has to
increase outward, i.e.\ with decreasing temperature. The sonic-point
conditions are thus limited to a curve in $\rho$--$T$ space for which
$\kappa_{\rm Ross}(\rho,T)=\kappa_{\rm Edd}$, or more precisely to the
branches of this curve with $\partial\kappa_{\rm Ross}/\partial T <
0$. This general picture has been confirmed by numerical wind models
from \citet{gra1:05,gra1:08}.

The wind condition Eq.\,\ref{eq:sonic} for the ratio between $P_{\rm
  rad}$ and $P_{\rm gas}$ at $R_{\rm s}$ is highly relevant in this
context. The condition that $\kappa_{\rm Ross}(\rho,T)=\kappa_{\rm
  Edd}$ can be mapped into the $P_{\rm rad}$--$P_{\rm gas}$ plane,
i.e.\ we have two conditions for $P_{\rm gas}$ and $P_{\rm rad}$ that
restrict the sonic-point conditions to the intersection points between
Eq.\,\ref{eq:sonic} and the condition that $\kappa_{\rm Ross}(P_{\rm
  gas},P_{\rm rad})=\kappa_{\rm Edd}$. On top of this
$\partial\kappa_{\rm Ross}/\partial T < 0$ has to be fulfilled, and
even more importantly, the sonic radius $R_{\rm s}$ of our wind
solution has to match the stellar radius. This last condition
over-determines the problem, i.e.\ we are in need of an additional
free parameter to obtain a solution.

A solution to this problem may be given by the envelope inflation
effect where the stellar radius becomes a function of the clumping
parameter $D$ in the sub-photospheric layers. In this picture a
compact WR star may not be able to launch a stationary stellar wind
because the sonic-point conditions cannot be fulfilled for the given
radius. The result may be a failed wind, i.e.\ material would be
accumulated in shells. Due to the higher density in these shells the
mean opacity would increase\footnote{\changedA note that this effect
  may be reduced by porosity cf.\ Sect.\ref{sec:clumping}} and lead to
an increase in the envelope inflation \citep[cf.][]{gra1:12}.

 At the moment where the stellar
radius enters a range for which a stationary wind solution can be
connected the system becomes quasi-stationary with a clumped
sub-photospheric structure that sustains a large stellar radius and a
wind solution that matches $\rho$, $T$, and $R$ of the inflated
envelope.

In Fig.\,\ref{fig:gamma} we visualize this scenario using hydrostatic
envelope solutions from \citet{gra1:12}. The stellar structure models
are computed for a mass of $14\,M_\sun$ ($\log(L/L_\odot)=5.43$), and
a composition with pure helium and Galactic metallicity of $Z=0.02$.
The envelope solutions follow a very similar relation with
$\kappa_{\rm Ross}(P_{\rm gas},P_{\rm rad})\approx \kappa_{\rm Edd}$
as imposed by the critical point conditions for optically thick winds.
The reason is that envelope solutions approaching the Eddington limit
cannot exceed it and thus have to stay near the Eddington limit. To
obtain a consistent envelope/wind solution we thus have to match the
envelope solution to our wind condition Eq.\,\ref{eq:sonic}.  At the
same time the radii of the envelope and wind solution have to match
each other at the connection point, and we have to fulfill
$\partial\kappa_{\rm Ross}/\partial T < 0$ which is equivalent to
$\partial\kappa_{\rm Ross}/\partial P_{\rm rad} < 0$.

Let us first discuss the envelope solution without clumping (i.e.\
with $D=1$) as indicated by the black curve in Fig.\,\ref{fig:gamma}.
Starting at high $P_{\rm gas}$ and $P_{\rm rad}$ (i.e.\ high $\rho$
and $T$) in the deep layers of the envelope the solution proceeds to
low $P_{\rm gas}$ and $P_{\rm rad}$ at the stellar surface. At
$\log(P_{\rm rad})\sim 6.2$ the solution has a `knee' where it goes to
low $P_{\rm gas}$ (i.e.\ low densities) and then returns to higher
$P_{\rm gas}$ through an inversion in gas pressure and density. This
is the location of the Fe-opacity peak. To avoid super-Eddington
opacities due to the increased opacity in this region the solution has
to move to low densities, i.e.\ it navigates {\em around} the
Fe-opacity peak.  During this excursion to low densities it crosses
the wind condition Eq.\,\ref{eq:sonic} (as indicated by the blue line)
two times.

The first intersection is located at high $P_{\rm rad}$ i.e.\ at
temperatures higher than the Fe-opacity peak ($\gtrsim 150$\,kK).  At
this point we have $\partial\kappa_{\rm Ross}/\partial P_{\rm rad} <
0$, i.e.\ a wind solution could be connected. The temperature
$T_\star$ of the corresponding wind solution lies slightly above
$100$\,kK (as indicated by the labels in Fig.\,\ref{fig:gamma}). The
second intersection at lower $P_{\rm rad}$ is located in the region of
the density inversion.  Here we have $\partial\kappa_{\rm
  Ross}/\partial P_{\rm rad} > 0$ and no wind solution is possible.

How do stellar radius and temperature of the envelope solution compare
to the stellar temperature ($T_\star \sim 100$\,kK) imposed by the
wind solution? For the un-clumped envelope solution with $D=1$ in
Fig.\,\ref{fig:gamma} the radius varies between $1.1\,R_\odot$ in the
deep layers and $1.2\,R_\odot$ at the stellar surface.  The (in this
case very moderate) inflation of the stellar envelope takes place near
the tip of the knee, i.e.\ at low densities. The first intersection
point is thus located {\em below} the inflated layers at $\sim
1.1\,R_\odot$.  For the given luminosity this corresponds to
$T_\star=125$\,kK, i.e.\ the radii of wind and envelope solution would
not fit together.  As we have already described above such a situation
may lead to a `failed' wind and radius inflation.

We note, however, that the possibility of a hot wind cannot be
strictly excluded on this basis. A lower $\dot{M}$ would lead to a
smaller $\tau_{\rm s}$, i.e.\ a hotter wind could be connected at
the same radius (cf.\ Sect.\,\ref{sec:uncertainties} and
Fig.\,\ref{fig:testmdot}).  In this case the star would appear hot
with a thin wind, i.e.\ it may reflect the appearance of a WO star
with a lower mass-loss rate than typical WC stars.

In case that such a connection cannot be found the star may be forced
to inflate. An inflated envelope solution with a clumping factor of
$D=16$ is indicated by the black dashed line in Fig.\,\ref{fig:gamma}.
For this solution the clumping factor has been set to $D=16$ for
temperatures below $200$\,kK and $D=1$ otherwise, with a smooth
transition between both regimes. This transition happens to occur just
in the region of the first intersection point. Due to the influence of
clumping the opacities shift to lower (mean) densities.  The envelope
solution is thus shifted towards lower values of $P_{\rm gas}$ in
Fig.\,\ref{fig:gamma}.

The lower density in the inflated zone leads to a stronger inflation
effect \citep[cf.][]{gra1:12}. Again, only the `knee' which is located
above the first intersection point is located within the inflated
zone. The first intersection point is thus still located at the same
radius as before ($\sim 1.1\,R_\odot$). Due to the increased envelope
inflation the layers above the knee are, however, shifted to much larger
radii of $7...8.5\,R_\sun$, corresponding to $T_\star=45...50$\,kK. As
we assume a constant clumping factor throughout the whole outer
envelope, also the surface layers above the knee are shifted to lower
densities. Due to this shift they almost coincide with the wind
condition Eq.\,\ref{eq:sonic} and it is easy to find a good connection
point with an optically thick wind solution with
$T_\star=45...50$\,kK. For this example the star would appear much
cooler than before, presumably as a WC\,8--9 subtype with a typical
mass-loss rate and clumping factor for this type of star.

For the present example we thus found two possible solutions that
correspond to the two cases discussed in Sect.\,\ref{sec:radii}.  1) a
solution with small radius, low mass-loss rate, and no
sub-photospheric clumping that may reflect the properties of WO stars,
and 2) a solution with large radius, standard mass-loss rate, and
sub-photospheric clumping that may reflect common late WC subtypes.
Notably, in our present approach the stellar radius is imposed by the
properties of the stellar wind, and sub-photospheric clumping may be
supported by the necessity to adjust the stellar radius accordingly.

\begin{figure}[]
  \parbox[b]{0.49\textwidth}{\center{\includegraphics[scale=0.4]{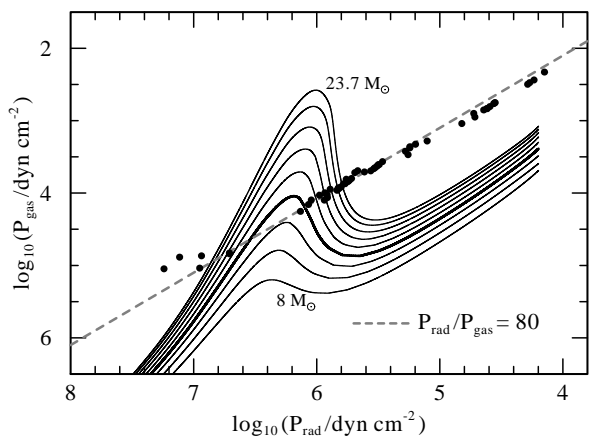}}}
  \caption{Comparison of envelope solutions in the mass range from
    $8\,M_\odot$ (bottom) to $23.7\,M_\odot$ (top) with the
    semi-empirical sonic-point conditions from
    Sect.\,\ref{sec:numeric} (black dots). The thick black curve
    indicates our previous example model with $14\,M_\odot$, and the
    grey dashed line the wind condition Eq.\,\ref{eq:sonic}.}
  \label{fig:comp}
\end{figure}

To explore the sonic-point conditions in the general case we compare
in Fig.\,\ref{fig:comp} envelope solutions without clumping for the
mass range from $8\,M_\odot$ to $23.7\,M_\odot$ with the
semi-empirical sonic-point conditions obtained in
Sect.\,\ref{sec:numeric}. With luminosities from $4.96\,L_\odot$ to
$5.81\,L_\odot$ the models largely cover the observed range of
luminosities for classical WR stars. The main differences with respect
to the previous example with $14\,M_\odot$ (indicated by the thick
black line) is the extension of the `knee' towards low densities. For
low masses the knee is almost absent, and there is no natural
intersection point with the wind condition from Eq.\,\ref{eq:sonic}.
For these objects one would always expect clumped sub-surface layers
as in case 2) to bring the envelope solution in agreement with the
wind condition Eq.\,\ref{eq:sonic}. For the highest masses the
situation is exactly the opposite.  There always exist two
intersection points, but due to the low densities at the tip of the
knee the un-clumped models are already inflated. Additional clumping
in the sub-surface layers would bring the envelopes above the
stability limit discussed by \citet{gra1:12}. This may prevent the
formation of clumped sub-surface layers for the more massive objects
and favour case 1). Generally, the shift between envelope solution and
wind condition is larger for lower masses, i.e.\ we expect larger
clumping factors in the sub-surface layers of lower mass stars.

The semi-empirical values from Sect.\,\ref{sec:numeric} are indicated
by black dots in Fig.\,\ref{fig:comp}. They fall in two groups that
are clearly separated, and most likely represent the cases 1) and 2)
discussed above.  The small group of objects with $\log{P_{\rm
    rad}}\approx 7$ consists of the three WO stars in our sample, plus
the only three galactic WC\,4 stars. While the WO stars clearly agree
with our case 1), the Galactic WC\,4 stars have higher mass-loss rates
but are still located in the expected temperature range. The rest of
the sample forms a sequence between $\log{P_{\rm rad}}\approx 4...6$,
largely in agreement with our case 2) with sub-surface clumping.
Problematic is that many objects (with $\log{P_{\rm rad}}\gtrsim 5.7$)
are located in the region of the density inversion where
$\partial\kappa_{\rm Ross}/\partial P_{\rm rad} > 0$.  Here it would
be necessary that the radial dependence of the clumping factor adjusts
in a way that the effective opacity increases outward, and a wind can
still be driven.

{\changedA As already discussed by \citet{gra1:12} the imposed clumping
  factors for case 2) reflect the values commonly observed in WR
  winds. The observed wind clumping may thus be initiated in the
  inflated sub-surface layers around the Fe-opacity peak. This would
  be in line with the work by \citet{can1:09} who suggested that the
  wind clumping in OB stars may be initiated in sub-surface convection
  zones caused by the same Fe-opacity peak. Their result was based on
  the observation of enhanced micro-turbulent velocities for OB stars
  with high convective velocities in their sub-surface layers. While
  our result is likely related to the one by \citeauthor{can1:09}, it
  is important to note that the nature of the instabilities in
  inflated envelopes is due to the dominance of radiation pressure,
  and thus markedly different from classical convection. The
  conditions discussed in our present work are only met in stars with
  high Eddington factors, and thus occur for much higher luminosities
  than the ones discussed by \citet{can1:09}.}

\section{Conclusions}
\label{sec:conclusions}

In this work we estimated sonic-point temperatures and densities for a
large sample of putatively single WC/WO stars in the Galaxy and LMC.
We found that they obey a relation with $P_{\rm rad}/P_{\rm gas}
\approx 80$ at the sonic point which is imposed by their optically
thick winds. This result is obtained with two methods, one of which
relies on detailed wind modelling while the other is based on a more
simplified model-independent approach. The result is based on the
assumption that the winds are optically thick and radiatively driven,
and is independent of effects such as wind clumping and porosity. Our
analysis of an artificial model sequence revealed that this `wind
condition' is ubiquitous for optically thick winds and only moderately
sensitive to detailed stellar parameters.

The high ratio of $P_{\rm rad}/P_{\rm gas}\approx80$ implies that
optically thick radiatively driven winds naturally emerge near the
Eddington limit.  Furthermore this wind condition imposes an outer
boundary condition on the stellar envelope that may considerably
affect its outer structure and radius.  We identify two possibilities
to connect the stellar envelopes of stars near the Eddington limit to
the imposed wind condition. In the first scenario the star is very
compact and the sonic-point temperature is higher than the temperature
of the Fe-opacity peak.  This scenario demands for relatively thin
winds, and may reflect the small group of WO stars with their
exceptionally hot cores and highly ionized winds, possibly also
including very early WC subtypes. In the second scenario the stellar
radius is large and the sonic point is located at lower temperatures
than the Fe-opacity peak.  In this case a solution is only possible if
the sub-photospheric layers at the location of the sonic point are
clumped.  This scenario is in line with the prediction of density
inhomogeneities due to strange modes in stellar envelopes near the
Eddington limit by \citet{gla1:08} and the enhanced envelope inflation
effect for clumped stellar envelopes by \citet{gra1:12}. The effect is
expected to be metallicity-dependent \citep{ish1:99}, and explains the
large radii observed for many WC stars at Galactic metallicity. The
imposed clumping factors reflect the values commonly observed in WR
winds, suggesting that the origin of WR wind clumping may be the
strange-mode instability in the inflated sub-surface layers.

An important new aspect is that the boundary conditions imposed by
optically thick winds may affect the radii of stars near the Eddington
limit. Towards the end of their lives massive stars tend to approach
the Eddington limit because their cores become chemically more and
more enriched, i.e.\ their mean molecular weight increases. The
effects discussed here may thus generally affect the radii and
effective temperatures of stars in these phases. In particular they
may be responsible for radius variations as observed for LBVs
\citep[cf.\ the discussion in][]{gra1:12}, and may lead to LBV-like
behaviour of some immediate supernova progenitors
\citep{kot1:06,gro1:13}.

\begin{acknowledgements}
  We thank the anonymous referee for his helpful comments, and STFC
  for financial support under grant No.\ ST/J001082/1, as well as the
  Department of Culture, Arts and Leisure in Northern Ireland.
\end{acknowledgements}

% BibTex Commands: 
%\bibliographystyle{aa}
%\bibliography{aamnem99,astro}

\appendix

\section{}

\begin{table*}
  \caption{Sonic-point conditions for WC/WO stars in the Galaxy and LMC. Empirical estimates from Sect.\,\ref{sec:rough}\label{tab:rough}}
  {\small \begin{tabular}{lr|rrrrrr|rrrrrrrr} \hline \hline \rule{0cm}{2.2ex}%
      name & ST & $\log(L)$ & $M$ & $\log(\dot{M})$ & $\varv_\infty$ & $\varv_{\rm esc}$ & $\eta$ & $R_{\rm ref}/R_\star$ & $f$ & $\tilde{\tau}_{\rm s}$ & $T_\star$ & $T_{\rm ref}$ & $T_{\rm s}$ & $\log(P_{\rm r})$ & $\log(P_{\rm g})$ \\
      \rule{0cm}{2.2ex}%
      &  & $[L_\odot]$ & $[M_\odot]$ & $[\frac{M_\odot}{{\rm yr}}]$ & $[\frac{{\rm km}}{{\rm s}}]$ & $[\frac{{\rm km}}{{\rm s}}]$ & & & & & [kK] & [kK] & [kK] & $[\frac{{\rm dyn}}{{\rm cm}^2}]$ & $[\frac{{\rm dyn}}{{\rm cm}^2}]$ \\
      \hline \rule{0cm}{2.2ex}%
      WR102 & 2 & 5.68 & 19.21 & -5.16 & 5000 & 3567 &  3.56 & 1.21 & 0.58 &   8.95 & 200.0 & 181.7 & 297.8 &  7.30 &  4.97 \\
      WR142 & 2 & 5.70 & 19.77 & -4.99 & 5000 & 3576 &  4.97 & 1.42 & 0.58 &  17.14 & 200.0 & 167.9 & 321.1 &  7.43 &  5.13 \\
      WR038 & 4 & 5.20 & 10.30 & -4.66 & 3200 & 2169 & 21.65 & 2.49 & 0.60 & 224.82 & 126.0 &  79.9 & 288.1 &  7.24 &  5.14 \\
      WR052 & 4 & 5.12 &  9.37 & -4.71 & 3225 & 1925 & 23.21 & 2.04 & 0.63 & 153.56 & 112.0 &  78.5 & 257.5 &  7.04 &  4.94 \\
      WR144 & 4 & 5.22 & 10.55 & -4.60 & 3500 & 1929 & 25.81 & 2.12 & 0.64 & 179.91 & 112.0 &  76.9 & 262.4 &  7.08 &  4.95 \\
      WR004 & 5 & 5.30 & 11.64 & -4.68 & 2528 & 1364 & 13.00 & 1.42 & 0.65 &  40.59 &  79.0 &  66.2 & 156.1 &  6.18 &  4.08 \\
      WR017 & 5 & 5.15 &  9.70 & -4.85 & 2231 & 1358 & 11.04 & 1.36 & 0.62 &  33.05 &  79.0 &  67.6 & 151.7 &  6.13 &  4.05 \\
      WR033 & 5 & 5.30 & 11.64 & -4.56 & 3342 & 1364 & 22.72 & 1.42 & 0.71 &  64.89 &  79.0 &  66.2 & 175.3 &  6.38 &  4.22 \\
      WR111 & 5 & 5.35 & 12.38 & -4.67 & 2398 & 1540 & 11.23 & 1.64 & 0.61 &  49.74 &  89.0 &  69.5 & 172.2 &  6.35 &  4.26 \\
      WR114 & 5 & 5.35 & 12.38 & -4.54 & 3200 & 1367 & 20.24 & 1.45 & 0.70 &  60.36 &  79.0 &  65.7 & 170.9 &  6.33 &  4.18 \\
      WR150 & 5 & 5.40 & 13.19 & -4.54 & 3000 & 1545 & 17.08 & 1.67 & 0.66 &  72.26 &  89.0 &  68.8 & 187.2 &  6.49 &  4.37 \\
      WR005 & 6 & 5.45 & 14.07 & -4.64 & 2120 & 1376 &  8.39 & 1.49 & 0.61 &  30.73 &  79.0 &  64.7 & 142.5 &  6.02 &  3.94 \\
      WR013 & 6 & 5.48 & 14.63 & -4.65 & 2000 & 1379 &  7.34 & 1.50 & 0.59 &  28.07 &  79.0 &  64.4 & 138.8 &  5.97 &  3.90 \\
      WR015 & 6 & 5.55 & 16.06 & -4.47 & 2675 & 1388 & 12.61 & 1.54 & 0.66 &  45.35 &  79.0 &  63.7 & 154.4 &  6.16 &  4.03 \\
      WR023 & 6 & 5.50 & 15.02 & -4.56 & 2342 & 1381 &  9.95 & 1.51 & 0.63 &  36.27 &  79.0 &  64.2 & 147.3 &  6.07 &  3.98 \\
      WR027 & 6 & 5.48 & 14.63 & -4.63 & 2100 & 1379 &  8.09 & 1.50 & 0.60 &  30.35 &  79.0 &  64.4 & 141.4 &  6.00 &  3.93 \\
      WR045 & 6 & 5.50 & 15.02 & -4.59 & 2200 & 1381 &  8.78 & 1.51 & 0.61 &  32.77 &  79.0 &  64.2 & 143.7 &  6.03 &  3.95 \\
      WR132 & 6 & 5.35 & 12.38 & -4.68 & 2400 & 1229 & 11.10 & 1.29 & 0.66 &  27.91 &  71.0 &  62.5 & 134.5 &  5.92 &  3.81 \\
      WR154 & 6 & 5.30 & 11.64 & -4.72 & 2300 & 1364 & 10.76 & 1.42 & 0.63 &  34.77 &  79.0 &  66.2 & 150.3 &  6.11 &  4.03 \\
      WR014 & 7 & 5.30 & 11.64 & -4.75 & 2194 & 1226 &  9.55 & 1.27 & 0.64 &  24.15 &  71.0 &  62.9 & 130.7 &  5.87 &  3.78 \\
      WR056 & 7 & 5.35 & 12.38 & -4.75 & 2009 & 1229 &  7.78 & 1.29 & 0.62 &  20.85 &  71.0 &  62.5 & 125.3 &  5.79 &  3.72 \\
      WR064 & 7 & 5.20 & 10.30 & -4.94 & 1700 & 1222 &  6.07 & 1.24 & 0.58 &  16.16 &  71.0 &  63.7 & 120.0 &  5.72 &  3.67 \\
      WR068 & 7 & 5.35 & 12.38 & -4.73 & 2100 & 1229 &  8.50 & 1.29 & 0.63 &  22.40 &  71.0 &  62.5 & 127.5 &  5.82 &  3.74 \\
      WR090 & 7 & 5.23 & 10.68 & -4.83 & 2053 & 1223 &  8.71 & 1.25 & 0.63 &  21.80 &  71.0 &  63.4 & 128.5 &  5.84 &  3.76 \\
      WR053 & 8 & 5.15 &  9.70 & -4.94 & 1800 &  859 &  7.12 & 1.04 & 0.68 &  11.32 &  50.0 &  49.1 &  85.0 &  5.12 &  3.03 \\
      WR057 & 8 & 5.30 & 11.64 & -4.84 & 1787 & 1088 &  6.42 & 1.17 & 0.62 &  14.15 &  63.0 &  58.2 & 106.3 &  5.51 &  3.44 \\
      WR060 & 8 & 5.40 & 13.19 & -4.65 & 2300 & 1093 & 10.04 & 1.19 & 0.68 &  21.10 &  63.0 &  57.7 & 115.9 &  5.66 &  3.55 \\
      WR135 & 8 & 5.28 & 11.35 & -4.82 & 1343 & 1087 &  5.18 & 1.34 & 0.55 &  16.73 &  63.0 &  54.5 & 103.6 &  5.46 &  3.47 \\
      WR059 & 9 & 4.90 &  7.29 & -5.13 & 1300 &  688 &  5.93 & 1.01 & 0.65 &   9.31 &  40.0 &  39.7 &  65.7 &  4.67 &  2.65 \\
      WR065 & 9 & 5.12 &  9.37 & -4.97 & 1300 &  687 &  5.22 & 1.02 & 0.65 &   8.35 &  40.0 &  39.6 &  63.8 &  4.62 &  2.59 \\
      WR069 & 9 & 5.20 & 10.30 & -4.98 & 1089 &  688 &  3.50 & 1.03 & 0.61 &   6.02 &  40.0 &  39.5 &  59.1 &  4.49 &  2.48 \\
      WR080 & 9 & 4.95 &  7.71 & -5.01 & 1600 &  773 &  8.66 & 1.04 & 0.67 &  13.79 &  45.0 &  44.2 &  80.2 &  5.02 &  2.98 \\
      WR081 & 9 & 5.15 &  9.70 & -4.71 & 1600 &  773 & 10.90 & 1.14 & 0.67 &  20.88 &  45.0 &  42.2 &  84.6 &  5.11 &  3.09 \\
      WR088 & 9 & 5.25 & 10.94 & -4.81 & 1500 &  689 &  6.45 & 1.03 & 0.69 &   9.99 &  40.0 &  39.4 &  66.3 &  4.69 &  2.63 \\
      WR092 & 9 & 5.22 & 10.55 & -4.81 & 1121 &  775 &  5.14 & 1.15 & 0.59 &  11.50 &  45.0 &  42.0 &  72.9 &  4.85 &  2.88 \\
      WR095 & 9 & 5.20 & 10.30 & -4.75 & 1900 &  774 & 10.57 & 1.06 & 0.71 &  16.65 &  45.0 &  43.8 &  83.1 &  5.08 &  2.99 \\
      WR103 & 9 & 5.14 &  9.59 & -4.84 & 1190 &  773 &  6.06 & 1.13 & 0.61 &  12.88 &  45.0 &  42.2 &  75.4 &  4.91 &  2.94 \\
      WR106 & 9 & 5.15 &  9.70 & -4.87 & 1100 &  773 &  5.15 & 1.14 & 0.59 &  11.33 &  45.0 &  42.2 &  73.1 &  4.86 &  2.89 \\
      WR117 & 9 & 5.35 & 12.38 & -4.45 & 2000 &  969 & 15.72 & 1.44 & 0.67 &  48.68 &  56.0 &  46.6 & 114.9 &  5.64 &  3.60 \\
      WR119 & 9 & 5.20 & 10.30 & -4.76 & 1300 &  774 &  6.99 & 1.15 & 0.63 &  14.65 &  45.0 &  42.0 &  77.4 &  4.96 &  2.96 \\
      WR121 & 9 & 5.20 & 10.30 & -4.83 & 1100 &  774 &  5.01 & 1.15 & 0.59 &  11.20 &  45.0 &  42.0 &  72.6 &  4.85 &  2.88 \\
      \hline \rule{0cm}{2.2ex}%
      WR011 & 8 & 5.00 &  8.30 & -5.00 & 1550 &  988 &  7.61 & 1.14 & 0.61 &  16.12 &  57.0 &  53.4 & 100.7 &  5.41 &  3.39 \\
      WR014 & 7 & 5.38 & 12.82 & -4.70 & 2055 & 1385 &  8.39 & 1.50 & 0.60 &  31.45 &  80.0 &  65.4 & 144.9 &  6.05 &  3.98 \\
      WR023 & 6 & 5.30 & 11.45 & -4.80 & 2280 & 1284 &  8.89 & 1.25 & 0.64 &  21.83 &  75.0 &  67.0 & 135.8 &  5.93 &  3.83 \\
      WR090 & 7 & 5.50 & 15.00 & -4.60 & 2045 & 1241 &  7.97 & 1.39 & 0.62 &  24.74 &  71.0 &  60.2 & 125.8 &  5.80 &  3.72 \\
      WR103 & 9 & 4.90 &  7.38 & -5.00 & 1140 &  831 &  7.04 & 1.17 & 0.58 &  16.54 &  48.0 &  44.5 &  84.3 &  5.10 &  3.16 \\
      WR111 & 5 & 5.30 & 11.33 & -4.80 & 2300 & 1551 &  8.97 & 1.50 & 0.60 &  33.87 &  91.0 &  74.3 & 167.5 &  6.30 &  4.22 \\
      WR135 & 8 & 5.20 & 10.50 & -4.90 & 1400 & 1094 &  5.46 & 1.27 & 0.56 &  15.58 &  63.0 &  56.0 & 104.6 &  5.48 &  3.47 \\
      WR154 & 6 & 5.02 &  8.23 & -5.00 & 2280 & 1365 & 10.69 & 1.25 & 0.63 &  26.64 &  80.0 &  71.6 & 152.3 &  6.13 &  4.05 \\
      WR111 & 5 & 5.45 & 13.99 & -4.90 & 2200 & 1476 &  4.82 & 1.20 & 0.60 &  11.58 &  85.0 &  77.6 & 135.2 &  5.93 &  3.80 \\
      \hline \rule{0cm}{2.2ex}%
      BR007 & 4 & 5.44 & 14.09 & -4.80 & 2500 & 1578 &  7.06 & 1.31 & 0.61 &  19.90 &  90.0 &  78.5 & 155.6 &  6.17 &  4.04 \\
      BR008 & 4 & 5.42 & 13.33 & -4.90 & 2300 & 1466 &  5.40 & 1.19 & 0.61 &  12.62 &  85.0 &  77.8 & 138.2 &  5.96 &  3.84 \\
      BR010 & 4 & 5.70 & 20.04 & -4.50 & 3000 & 1620 &  9.29 & 1.46 & 0.65 &  30.34 &  90.0 &  74.6 & 163.8 &  6.26 &  4.09 \\
      BR043 & 4 & 5.65 & 18.24 & -4.50 & 2900 & 1591 & 10.08 & 1.53 & 0.65 &  36.56 &  90.0 &  72.7 & 167.2 &  6.29 &  4.15 \\
      BR050 & 4 & 5.68 & 19.69 & -4.40 & 3200 & 1534 & 13.06 & 1.52 & 0.68 &  44.43 &  85.0 &  69.0 & 166.5 &  6.29 &  4.12 \\
      BR074 & 4 & 5.44 & 13.71 & -4.80 & 2600 & 1470 &  7.34 & 1.23 & 0.64 &  17.46 &  85.0 &  76.6 & 147.0 &  6.07 &  3.93 \\
      SND2 & 2 & 5.30 & 11.12 & -5.00 & 4100 & 2533 & 10.09 & 1.47 & 0.62 &  35.28 & 150.0 & 123.7 & 281.9 &  7.20 &  5.00 \\
      \hline \end{tabular}}
  \tablefoot{Columns 1--8 indicate names, spectral subtypes, and empirical stellar parameters for putatively single WC/WO stars compiled from various sources in the literature (top: Galactic WC/WO stars from \citet{san1:12}; middle: Galactic WC stars from \citet{hil1:99,dem1:00,des1:00,sma1:01,cro2:06,gra1:02}; bottom: LMC WC/WO stars from \citet{cro1:02,cro1:00}). {\changedA Typical error margins given by \citet{san1:12} are $\Delta \log(L)=\pm 0.3;$ $\Delta \log(\dot{M}_{\rm t}) = \Delta \log(\dot{M} \sqrt{D}/(\varv_\infty L^{3/4}))= \pm 0.15$; $\Delta \log(T_\star)=\pm 0.05$. A clumping factor $D=10$ has been adopted for all objects.} Columns 9-16 indicate wind parameters that are directly deduced from the empirical stellar parameters, as outlined in Sect.\,\ref{sec:rough}. $P_{\rm r}$ and $P_{\rm g}$ denote the radiation and gas pressure at the sonic point obtained from this method.}
\end{table*} 

\begin{table*}
  \caption{Sonic-point conditions for WC/WO stars in the Galaxy and LMC. Numerical estimates from Sect.\,\ref{sec:numeric}\label{tab:numeric}}
  {\small \begin{tabular}{lr|rrrrrr|rrrrrrrr} \hline \hline \rule{0cm}{2.2ex}%
      name & ST & $\log(L)$ & $M$ & $\log(\dot{M})$ & $\varv_\infty$ & $\varv_{\rm esc}$ & $\eta$ & $R_{\rm ref}/R_\star$ & $f$ & $\tilde{\tau}_{\rm s}$ & $T_\star$ & $T_{\rm ref}$ & $T_{\rm s}$ & $\log(P_{\rm r})$ & $\log(P_{\rm g})$ \\
      \rule{0cm}{2.2ex}%
      &  & $[L_\odot]$ & $[M_\odot]$ & $[\frac{M_\odot}{{\rm yr}}]$ & $[\frac{{\rm km}}{{\rm s}}]$ & $[\frac{{\rm km}}{{\rm s}}]$ & & & & & [kK] & [kK] & [kK] & $[\frac{{\rm dyn}}{{\rm cm}^2}]$ & $[\frac{{\rm dyn}}{{\rm cm}^2}]$ \\
      \hline \rule{0cm}{2.2ex}%
      WR102 & 2 & 5.68 & 19.21 & -5.16 & 5000 & 3567 &  3.56 &  2.56 & 0.43 &   8.31 & 200.0 & 124.5 & 268.0 &  7.11 &  4.89 \\
      WR142 & 2 & 5.70 & 19.77 & -4.99 & 5000 & 3576 &  4.97 &  3.51 & 0.43 &  11.55 & 200.0 & 105.7 & 288.3 &  7.24 &  5.05 \\
      WR038 & 4 & 5.20 & 10.30 & -4.66 & 3200 & 2169 & 21.65 & 16.37 & 0.47 &  46.57 & 126.0 &  29.9 & 243.9 &  6.95 &  5.04 \\
      WR052 & 4 & 5.12 &  9.37 & -4.71 & 3225 & 1925 & 23.21 & 16.53 & 0.52 &  44.63 & 112.0 &  26.4 & 212.3 &  6.71 &  4.83 \\
      WR144 & 4 & 5.22 & 10.55 & -4.60 & 3500 & 1929 & 25.81 & 18.34 & 0.55 &  47.15 & 112.0 &  24.9 & 212.1 &  6.71 &  4.84 \\
      WR004 & 5 & 5.30 & 11.64 & -4.68 & 2528 & 1364 & 13.00 &  8.27 & 0.59 &  22.16 &  79.0 &  27.1 & 128.4 &  5.84 &  3.96 \\
      WR017 & 5 & 5.15 &  9.70 & -4.85 & 2231 & 1358 & 11.04 &  7.26 & 0.54 &  20.52 &  79.0 &  29.0 & 127.4 &  5.82 &  3.94 \\
      WR033 & 5 & 5.30 & 11.64 & -4.56 & 3342 & 1364 & 22.72 & 14.74 & 0.68 &  33.59 &  79.0 &  19.9 & 136.2 &  5.94 &  4.10 \\
      WR111 & 5 & 5.35 & 12.38 & -4.67 & 2398 & 1540 & 11.23 &  7.55 & 0.51 &  22.00 &  89.0 &  32.0 & 146.7 &  6.07 &  4.16 \\
      WR114 & 5 & 5.35 & 12.38 & -4.54 & 3200 & 1367 & 20.24 & 12.98 & 0.67 &  30.43 &  79.0 &  21.3 & 134.2 &  5.91 &  4.06 \\
      WR150 & 5 & 5.40 & 13.19 & -4.54 & 3000 & 1545 & 17.08 & 11.06 & 0.60 &  28.67 &  89.0 &  26.1 & 152.3 &  6.13 &  4.25 \\
      WR005 & 6 & 5.45 & 14.07 & -4.64 & 2120 & 1376 &  8.39 &  5.53 & 0.51 &  16.41 &  79.0 &  33.3 & 121.7 &  5.74 &  3.83 \\
      WR013 & 6 & 5.48 & 14.63 & -4.65 & 2000 & 1379 &  7.34 &  4.94 & 0.49 &  15.12 &  79.0 &  35.3 & 119.9 &  5.72 &  3.80 \\
      WR015 & 6 & 5.55 & 16.06 & -4.47 & 2675 & 1388 & 12.61 &  7.92 & 0.60 &  20.97 &  79.0 &  27.7 & 126.4 &  5.81 &  3.92 \\
      WR023 & 6 & 5.50 & 15.02 & -4.56 & 2342 & 1381 &  9.95 &  6.40 & 0.55 &  18.05 &  79.0 &  30.9 & 123.5 &  5.77 &  3.87 \\
      WR027 & 6 & 5.48 & 14.63 & -4.63 & 2100 & 1379 &  8.09 &  5.36 & 0.51 &  15.98 &  79.0 &  33.9 & 121.1 &  5.73 &  3.82 \\
      WR045 & 6 & 5.50 & 15.02 & -4.59 & 2200 & 1381 &  8.78 &  5.72 & 0.53 &  16.72 &  79.0 &  32.8 & 122.0 &  5.75 &  3.84 \\
      WR132 & 6 & 5.35 & 12.38 & -4.68 & 2400 & 1229 & 11.10 &  6.86 & 0.62 &  18.12 &  71.0 &  26.8 & 109.6 &  5.56 &  3.69 \\
      WR154 & 6 & 5.30 & 11.64 & -4.72 & 2300 & 1364 & 10.76 &  6.93 & 0.55 &  19.60 &  79.0 &  29.7 & 125.9 &  5.80 &  3.92 \\
      WR014 & 7 & 5.30 & 11.64 & -4.75 & 2194 & 1226 &  9.55 &  6.02 & 0.58 &  16.51 &  71.0 &  28.7 & 108.1 &  5.54 &  3.66 \\
      WR056 & 7 & 5.35 & 12.38 & -4.75 & 2009 & 1229 &  7.78 &  5.02 & 0.54 &  14.38 &  71.0 &  31.5 & 105.4 &  5.49 &  3.61 \\
      WR064 & 7 & 5.20 & 10.30 & -4.94 & 1700 & 1222 &  6.07 &  4.19 & 0.47 &  12.86 &  71.0 &  34.6 & 103.7 &  5.47 &  3.57 \\
      WR068 & 7 & 5.35 & 12.38 & -4.73 & 2100 & 1229 &  8.50 &  5.40 & 0.56 &  15.20 &  71.0 &  30.3 & 106.4 &  5.51 &  3.63 \\
      WR090 & 7 & 5.23 & 10.68 & -4.83 & 2053 & 1223 &  8.71 &  5.56 & 0.56 &  15.77 &  71.0 &  29.9 & 107.4 &  5.53 &  3.65 \\
      WR053 & 8 & 5.15 &  9.70 & -4.94 & 1800 &  859 &  7.12 &  4.32 & 0.66 &  10.89 &  50.0 &  24.0 &  67.7 &  4.72 &  2.90 \\
      WR057 & 8 & 5.30 & 11.64 & -4.84 & 1787 & 1088 &  6.42 &  4.14 & 0.55 &  11.73 &  63.0 &  30.8 &  88.9 &  5.20 &  3.33 \\
      WR060 & 8 & 5.40 & 13.19 & -4.65 & 2300 & 1093 & 10.04 &  6.10 & 0.65 &  15.54 &  63.0 &  25.3 &  93.2 &  5.28 &  3.42 \\
      WR135 & 8 & 5.28 & 11.35 & -4.82 & 1343 & 1087 &  5.18 &  3.80 & 0.43 &  12.15 &  63.0 &  32.2 &  91.3 &  5.24 &  3.36 \\
      WR059 & 9 & 4.90 &  7.29 & -5.13 & 1300 &  688 &  5.93 &  3.72 & 0.62 &   9.55 &  40.0 &  20.7 &  52.6 &  4.29 &  2.50 \\
      WR065 & 9 & 5.12 &  9.37 & -4.97 & 1300 &  687 &  5.22 &  3.28 & 0.62 &   8.40 &  40.0 &  22.0 &  51.1 &  4.23 &  2.44 \\
      WR069 & 9 & 5.20 & 10.30 & -4.98 & 1089 &  688 &  3.50 &  2.39 & 0.55 &   6.41 &  40.0 &  25.9 &  48.6 &  4.15 &  2.33 \\
      WR080 & 9 & 4.95 &  7.71 & -5.01 & 1600 &  773 &  8.66 &  5.26 & 0.66 &  13.25 &  45.0 &  19.5 &  63.7 &  4.62 &  2.83 \\
      WR081 & 9 & 5.15 &  9.70 & -4.71 & 1600 &  773 & 10.90 &  6.64 & 0.66 &  16.64 &  45.0 &  17.3 &  67.3 &  4.71 &  2.95 \\
      WR088 & 9 & 5.25 & 10.94 & -4.81 & 1500 &  689 &  6.45 &  3.90 & 0.68 &   9.54 &  40.0 &  20.2 &  52.1 &  4.27 &  2.47 \\
      WR092 & 9 & 5.22 & 10.55 & -4.81 & 1121 &  775 &  5.14 &  3.51 & 0.51 &  10.13 &  45.0 &  24.0 &  61.4 &  4.55 &  2.75 \\
      WR095 & 9 & 5.20 & 10.30 & -4.75 & 1900 &  774 & 10.57 &  6.24 & 0.72 &  14.80 &  45.0 &  17.9 &  64.6 &  4.64 &  2.85 \\
      WR103 & 9 & 5.14 &  9.59 & -4.84 & 1190 &  773 &  6.06 &  4.03 & 0.54 &  11.35 &  45.0 &  22.4 &  62.8 &  4.59 &  2.81 \\
      WR106 & 9 & 5.15 &  9.70 & -4.87 & 1100 &  773 &  5.15 &  3.54 & 0.50 &  10.31 &  45.0 &  23.9 &  61.7 &  4.56 &  2.76 \\
      WR117 & 9 & 5.35 & 12.38 & -4.45 & 2000 &  969 & 15.72 &  9.72 & 0.65 &  24.44 &  56.0 &  17.7 &  92.1 &  5.26 &  3.47 \\
      WR119 & 9 & 5.20 & 10.30 & -4.76 & 1300 &  774 &  6.99 &  4.48 & 0.57 &  12.23 &  45.0 &  21.2 &  63.6 &  4.61 &  2.83 \\
      WR121 & 9 & 5.20 & 10.30 & -4.83 & 1100 &  774 &  5.01 &  3.45 & 0.50 &  10.02 &  45.0 &  24.2 &  61.3 &  4.55 &  2.75 \\
      \hline \rule{0cm}{2.2ex}%                               
      WR011 & 8 & 5.00 &  8.30 & -5.00 & 1550 &  988 &  7.61 &  4.97 & 0.54 &  14.25 &  57.0 &  25.5 &  84.2 &  5.10 &  3.28 \\
      WR014 & 7 & 5.38 & 12.82 & -4.70 & 2055 & 1385 &  8.39 &  5.60 & 0.50 &  16.95 &  80.0 &  33.5 & 124.6 &  5.78 &  3.87 \\
      WR023 & 6 & 5.30 & 11.45 & -4.80 & 2280 & 1284 &  8.89 &  5.60 & 0.57 &  15.56 &  75.0 &  31.4 & 112.9 &  5.61 &  3.71 \\
      WR090 & 7 & 5.50 & 15.00 & -4.60 & 2045 & 1241 &  7.97 &  5.13 & 0.55 &  14.66 &  71.0 &  31.2 & 105.9 &  5.50 &  3.61 \\
      WR103 & 9 & 4.90 &  7.38 & -5.00 & 1140 &  831 &  7.04 &  4.86 & 0.49 &  14.55 &  48.0 &  21.7 &  71.6 &  4.82 &  3.04 \\
      WR111 & 5 & 5.30 & 11.33 & -4.80 & 2300 & 1551 &  8.97 &  6.08 & 0.49 &  18.39 &  91.0 &  36.6 & 144.8 &  6.04 &  4.10 \\
      WR135 & 8 & 5.20 & 10.50 & -4.90 & 1400 & 1094 &  5.46 &  3.93 & 0.44 &  12.33 &  63.0 &  31.7 &  91.3 &  5.24 &  3.38 \\
      WR154 & 6 & 5.02 &  8.23 & -5.00 & 2280 & 1365 & 10.69 &  6.98 & 0.55 &  19.69 &  80.0 &  30.0 & 127.6 &  5.83 &  3.93 \\
      WR111 & 5 & 5.45 & 13.99 & -4.90 & 2200 & 1476 &  4.82 &  3.27 & 0.49 &   9.83 &  85.0 &  46.9 & 116.5 &  5.67 &  3.69 \\
      \hline \rule{0cm}{2.2ex}%                               
      BR007 & 4 & 5.44 & 14.09 & -4.80 & 2500 & 1578 &  7.06 &  4.67 & 0.52 &  13.63 &  90.0 &  41.3 & 132.5 &  5.89 &  3.95 \\
      BR008 & 4 & 5.42 & 13.33 & -4.90 & 2300 & 1466 &  5.40 &  3.57 & 0.51 &  10.56 &  85.0 &  44.8 & 118.1 &  5.69 &  3.72 \\
      BR010 & 4 & 5.70 & 20.04 & -4.50 & 3000 & 1620 &  9.29 &  5.85 & 0.58 &  15.98 &  90.0 &  36.8 & 135.8 &  5.93 &  4.00 \\
      BR043 & 4 & 5.65 & 18.24 & -4.50 & 2900 & 1591 & 10.08 &  6.38 & 0.58 &  17.58 &  90.0 &  35.2 & 139.3 &  5.98 &  4.03 \\
      BR050 & 4 & 5.68 & 19.69 & -4.40 & 3200 & 1534 & 13.06 &  8.14 & 0.63 &  20.78 &  85.0 &  29.3 & 134.4 &  5.92 &  4.02 \\
      BR074 & 4 & 5.44 & 13.71 & -4.80 & 2600 & 1470 &  7.34 &  4.67 & 0.57 &  13.03 &  85.0 &  39.1 & 122.9 &  5.76 &  3.81 \\
      SND2 & 2 & 5.30 & 11.12 & -5.00 & 4100 & 2533 & 10.09 &  6.79 & 0.50 &  20.19 & 150.0 &  56.5 & 241.9 &  6.94 &  4.87 \\
      \hline \rule{0cm}{2.2ex}%
      hydro &   & 5.45 & 13.63 & -5.14 & 2020 & 2397 &  2.56 &  2.28 & 0.26 &   9.81 & 140.0 &  92.5 & 195.2 &  6.56 &  4.41 \\
      beta  &   & 5.45 & 13.64 & -5.14 & 2010 & 2397 &  2.53 &  2.61 & 0.25 &  10.20 & 140.0 &  86.5 & 201.3 &  6.62 &  4.42 \\
      \hline \end{tabular}}
  \tablefoot{Columns 1--8 indicate names, spectral subtypes, and empirical stellar parameters for putatively single WC/WO stars in analogy to Tab\,\ref{tab:rough}. The two bottom rows indicate wind models for WC stars from \citet{gra1:05}.
    Columns 9-16 indicate wind parameters that are numerically determined, as outlined in Sect.\,\ref{sec:numeric}. $P_{\rm r}$ and $P_{\rm g}$ denote the radiation and gas pressure at the sonic point obtained from this method.}
\end{table*} 

\end{document}